\documentclass[sigplan,nonacm]{acmart}

\usepackage[english]{babel}
\usepackage{blindtext}
\usepackage{algorithm}
\usepackage{algpseudocode}
\usepackage{subcaption}
\usepackage{float}
\usepackage{lipsum}
\usepackage{multirow}

\def\shownotes{1}       

\ifnum\shownotes=1
\newcommand{\authnote}[2]{{ $\ll$\textsf{\footnotesize #1 notes: #2}$\gg$}}
\else
\newcommand{\authnote}[2]{}
\fi

\providecommand{\ie}{\emph{i.e.,} }
\providecommand{\eg}{\emph{e.g.,} }

\providecommand{\etc}{\emph{etc.}}      

\providecommand{\myparab}[1]{\vspace{1pt}\noindent\textbf{#1} }

\providecommand{\sysname}{\textsc{Morphlux}\xspace}
\providecommand{\ipronics}{\textsc{iPronics}\xspace}
\providecommand{\orchestrator}{\textsc{MorphMgr}\xspace}
\providecommand{\allocator}{\textit{allocator}\xspace}
\providecommand{\faultmanager}{\textit{fault manager}\xspace}
\providecommand{\controlplane}{\textit{hardware control plane}\xspace}

\providecommand{\passage}{\textsc{Morphlux}\xspace}
\providecommand{\alltoall}{\textsc{AllToAll}\xspace}
\providecommand{\allreduce}{\textsc{AllReduce}\xspace}
\providecommand{\allgather}{\textsc{AllGather}\xspace}
\providecommand{\reducescatter}{\textsc{ReduceScatter}\xspace}
\providecommand{\ipronics}{\textsc{iPronics}\xspace}

\expandafter\def\expandafter\normalsize\expandafter{%
        \normalsize%
        \setlength\abovedisplayskip{1pt}%
        \setlength\belowdisplayskip{2pt}%
        \setlength\abovedisplayshortskip{-2pt}%
        \setlength\belowdisplayshortskip{2pt}%
}
\usepackage{enumitem}
\newlist{compactitem}{itemize}{1}
\setlist[compactitem,1]{label=\textbullet, left=0pt, itemsep=1pt, topsep=1pt, parsep=0pt, partopsep=0pt}
\usepackage[]{hyperref}
\hypersetup{
    draft=false,            
    unicode=false,          
    pdftoolbar=true,        
    pdfmenubar=true,        
    pdffitwindow=false,     
    pdftitle={},            
    pdfauthor={}
    pdfsubject={},          
    pdfnewwindow=true,      
    pdfkeywords={keywords}, 
    colorlinks=true,        
    linkcolor=black,      
    citecolor=black,    
    filecolor=black,      
    urlcolor=blue,      
}
\usepackage[capitalise]{cleveref}

\begin{document}
\title{\LARGE{\sysname: Transforming Torus Fabrics for Efficient Multi-tenant ML}}


\author{Abhishek Vijaya Kumar}
\affiliation{
  \institution{Cornell University}
  \city{Ithaca}
  \state{NY}
  \country{USA}
}
\author{Eric Ding}
\affiliation{
  \institution{Cornell University}
  \city{Ithaca}
  \state{NY}
  \country{USA}
}
\author{Arjun Devraj}
\affiliation{
  \institution{Cornell University}
  \city{Ithaca}
  \state{NY}
  \country{USA}
}

\author{Darius Bunandar}
\affiliation{
  \institution{Lightmatter}
  \city{Boston}
  \state{MA}
  \country{USA}
}

\author{Rachee Singh}
\affiliation{
  \institution{Cornell University}
  \city{Ithaca}
  \state{NY}
  \country{USA}
}

\begin{abstract}
We develop \sysname, a server-scale programmable photonic fabric to interconnect accelerators within servers. We show that augmenting state-of-the-art torus-based ML datacenters with \sysname can improve the bandwidth of tenant compute allocations by up to 66\%, reduce compute fragmentation by up to 70\%, and minimize the blast radius of chip failures. We develop a novel end-to-end hardware prototype of \sysname to demonstrate these performance benefits which translate to $1.72\times$ improvement in training throughput of ML models. By rapidly programming the server-scale fabric in our hardware testbed, \sysname can  replace a failed accelerator chip with a healthy one in 1.2 seconds~\footnote{Our code and data are available at: \url{https://github.com/morphlux/morphlux} \label{footnote:code}}. 

\end{abstract}

\maketitle
\title{}
\section{Introduction}
\label{sec:intro}

Collective communication primitives (\eg \allreduce, \alltoall) lie on the critical path of distributed ML workloads. As a result, large cloud providers have developed specialized network fabrics that improve the efficiency of collective communication during distributed training of large models~\cite{googletpuv4,aws_ultraclusters_2024,aws_trn2_blog_2024,aws_neuron_trn1_arch_2024}. For instance, Google and Amazon have deployed datacenter fabrics that connect accelerators in 3D-torus topology~\cite{googletpuv4,tpu-resilience,lightwave,aws_ultraclusters_2024,aws_trn2_blog_2024,aws_neuron_trn1_arch_2024}. When paired with multi-dimensional ring algorithms~\cite{bucket-ring,nd_torus}, torus-shaped accelerator topologies can achieve contention-free collective communication in training jobs spanning thousands of accelerators.

\myparab{Multi-tenancy crisis of tori.}
Torus-based datacenter architectures delivered on their promise for training massive ML models that require entire datacenter-scale allocations for weeks or months~\cite{lightwave,tpu-resilience}. However, the ML landscape is shifting from monolithic training to a diverse ecosystem of smaller inference and fine-tuning workloads. As foundation models mature, most cloud users need compute for inference on pre-trained models or fine-tuning with domain-specific data---tasks requiring far fewer resources than the original training jobs. We show that there is a critical mismatch: multiple inference jobs must coexist in ML datacenters, but torus-based fabrics are optimized for single, large-scale training jobs, limiting effective multi-tenancy---the ability to allocate arbitrary-sized compute subsets while maintaining consistent network performance---due to three challenges:

\begin{compactitem}
    \item \textbf{Idle bandwidth.} The egress bandwidth from an accelerator in torus topologies is statically partitioned along the three torus dimensions. As a result, if a tenant slice is smaller than a full torus-shaped rack, accelerators in that slice can only use a fraction of their egress bandwidth, leaving the rest idle. This degrades the performance of collective communication and end-to-end ML training throughput. Our experiments on the torus-based fabric of a large cloud provider confirm that bandwidth available to accelerators in smaller tenant slices can be up to 66\% lower than the bandwidth available in full-rack slices.

    \item \textbf{Compute fragmentation.} Torus-shaped datacenter fabrics are susceptible to compute fragmentation: a phenomenon where an existing allocation of compute resources to tenants makes it impossible to allocate topologies of remaining resources to new tenants. These limitations hinder the flexibility needed to support many small inference tasks. We develop a simulator for the torus-based fabric of a large cloud provider and show that compute fragmentation can be as high as 70\% at scale.

    \item \textbf{Large blast radius of accelerator failures.} Failure of even one accelerator requires migrating an ML job to an entirely new rack such that torus shape constraints continue to be met, which is costly and impractical at scale.

\end{compactitem}

\begin{figure}[h!]
\includegraphics[width=0.33\textwidth]{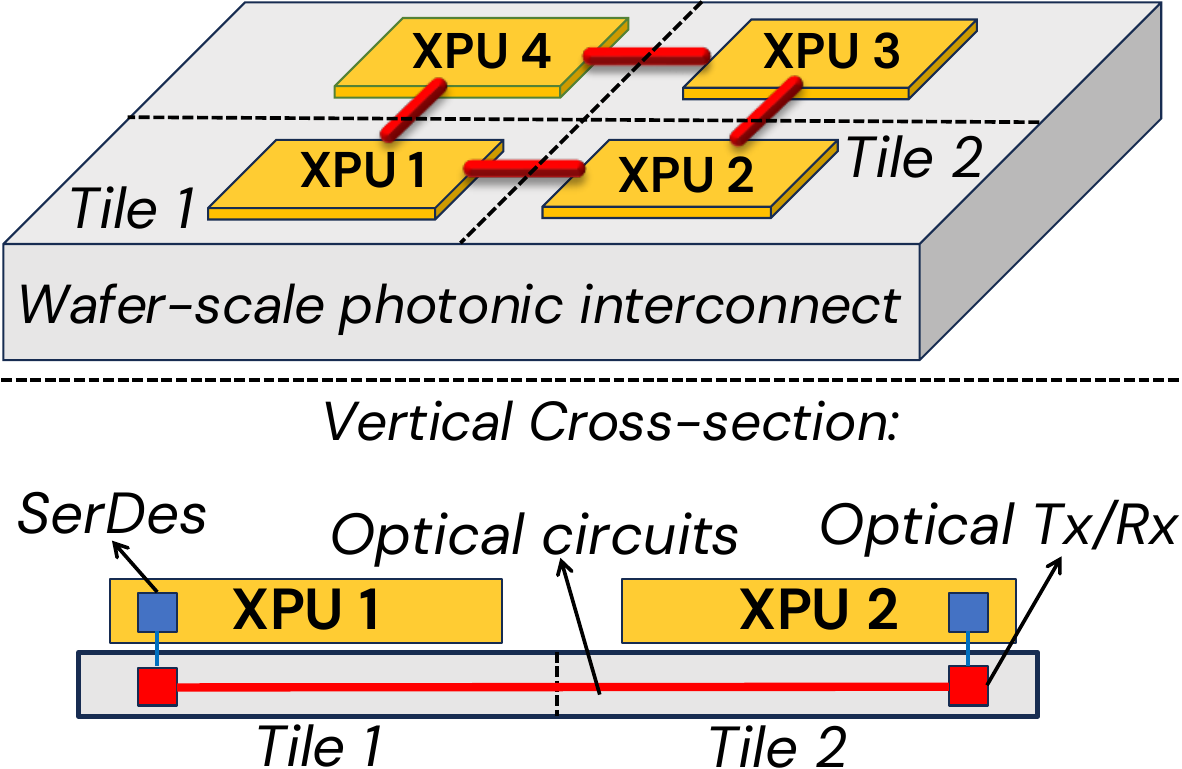}
      \caption{\small{shows how ML accelerators (XPUs) are stacked on top of the optical interposer in a single server. Connections between accelerators are direct optical links. Bottom figure shows a vertical cross-section of the server where two 2 XPUs connect on two tiles and communicate via the underlying optical link (shown in red). The connections between accelerators can be programmed to redirect egress bandwidth from one accelerator to another.}}
      \label{fig:wafer}
 \end{figure} 

\myparab{Bridging flexibility and performance.}
While traditional packet-switched datacenter fabrics can allocate any size of compute to tenants, they suffer from network contention that prevents efficient collective communication~\cite{cassini,taccl}. On the other hand, torus-based datacenter fabrics enable contention-free connectivity for optimal collective communication but they constrain compute allocation, restricting multi-tenancy. In this work we address this lack of flexibility in torus-based datacenter fabrics while preserving their network performance characteristics. Our key idea is enabling each accelerator in the torus-based datacenter to dynamically adjust their connectivity based on workload demands. We show that this capability addresses the limitations of torus-based fabrics in multi-tenant environments.

\myparab{\sysname: programmable optics in multi-accelerator servers.}
To achieve this vision, we connect accelerators within a server using programmable optical links instead of traditional electrical ones. In our design, \sysname, accelerators are stacked on top of a wafer-scale optical interposer~\cite{hotchips34}. Silicon-based optical waveguides etched on the wafer provide photonic connectivity between these accelerators (Figure~\ref{fig:wafer}). We note that optical interposer-based designs have recently seen commercial viability, independent of our proposal~\cite{hotchips34}. However, we leverage them to tackle the challenges of multi-tenancy in torus-based ML datacenters. \sysname redirects egress bandwidth from an accelerator along connections where it is needed instead of statically partitioning it among pre-defined inter-accelerator links. This allows accelerators to send data at their full egress bandwidth to a changing set of accelerator neighbors based on the size of tenant allocation and accelerator failures without sacrificing the benefits of contention-free direct connectivity of tori. \sysname achieves three key benefits:

\begin{compactitem}
    \item \myparab{Eliminates idle bandwidth.} 
    \sysname redirects bandwidth from an accelerator along connections within a tenant slice to achieve full egress bandwidth regardless of slice size, achieving up to 66\% higher bandwidth and up to 2X better end-to-end ML training throughput.

    \item \myparab{Eliminates compute fragmentation.} \sysname configures optical switches on the interposer to establish direct-connectivity between non-contiguous compute resources, reducing compute fragmentation by up to 70\% at scale.

    \item \myparab{Minimizes blast radius of accelerator failures.} 
    \sysname routes around accelerator failures, without causing network contention in other tenant slices, limiting the blast radius of a chip failure to only the server with the failed chip. Our end-to-end hardware experiments show that \sysname can recover from accelerator failures in roughly 1 second, which is magnitudes faster than the alternative of migrating entire jobs~\cite{tpu-resilience}.
\end{compactitem}

\myparab{Infrequent adaptation of photonic fabric.}
Creating optical circuits on \sysname incurs a delay of several microseconds~\cite{hotchips34}, thus \sysname creates new circuits in only two scenarios (1) at the time of allocating a slice of compute to tenants and (2) to handle sudden accelerator failures. Both these scenarios are tolerant to the microsecond-scale delay incurred in configuring optical links in \sysname.

\subsection{Overcoming Semiconductor Packaging Barriers}
Building a hardware demonstration of chip-to-chip photonic interconnects like \sysname faces a fundamental economic barrier: implementing true semiconductor packaging, where GPUs are directly bonded onto a silicon photonic interposer, requires access to advanced fabrication facilities and would cost tens of millions of dollars for even a single prototype. To overcome this challenge while still validating \sysname's core innovations, we construct an alternative prototype that preserves the critical silicon photonic mesh technology of the interposer but connects GPUs via optical fibers rather than direct chip bonding. This approach allows us to demonstrate \sysname's programmable topology capabilities using a testbed consisting of an off-the-shelf silicon photonic mesh connected to four GPU servers through optical transceivers.

Despite the fiber coupling introducing significant optical losses that limit our demonstration to 10 Gbps connections---far below the hundreds of Gbps achievable with direct chip integration---our prototype validates \sysname's key systems contributions. We demonstrate dynamic bandwidth reallocation that eliminates bandwidth underutilization in static torus topologies, achieve fast GPU failure recovery (1.2 seconds) through in-place topology reconfiguration. Most importantly, \sysname's system-level benefits are orthogonal to the raw link speed; the same programmability mechanisms that improve resource utilization at 10 Gbps will deliver proportionally greater absolute bandwidth gains when deployed with production-speed interconnects. Our prototype thus validates \sysname's architectural innovations while acknowledging that full performance realization awaits the economic feasibility of production-scale semiconductor packaging.

\myparab{Open-source artifacts and discussion.}
We have released \sysname's code, data from simulation and hardware experiments presented in the paper $^{\ref{footnote:code}}$. We note that there are practical challenges that limit large-scale adoption of silicon photonics and advanced semiconductor packaging technology that \sysname relies on. These concerns include thermal stabilization of ring resonators, laser power constraints, heat dissipation concerns~\cite{cpo}. Research across computer science, engineering, and engineering physics is actively tackling these challenges~\cite{mrr-stability1, mrr-stability2}. Our work takes a step forward by showing the feasibility of \sysname and the end-to-end benefits of server-scale photonic fabrics for ML.

\section{Torus-based datacenter fabrics for ML}
\label{sec:tpu_fabric}

\begin{figure*} 
    \centering
    \begin{subfigure}[b]{0.28\textwidth}
      \centering
      \includegraphics[width=\textwidth]{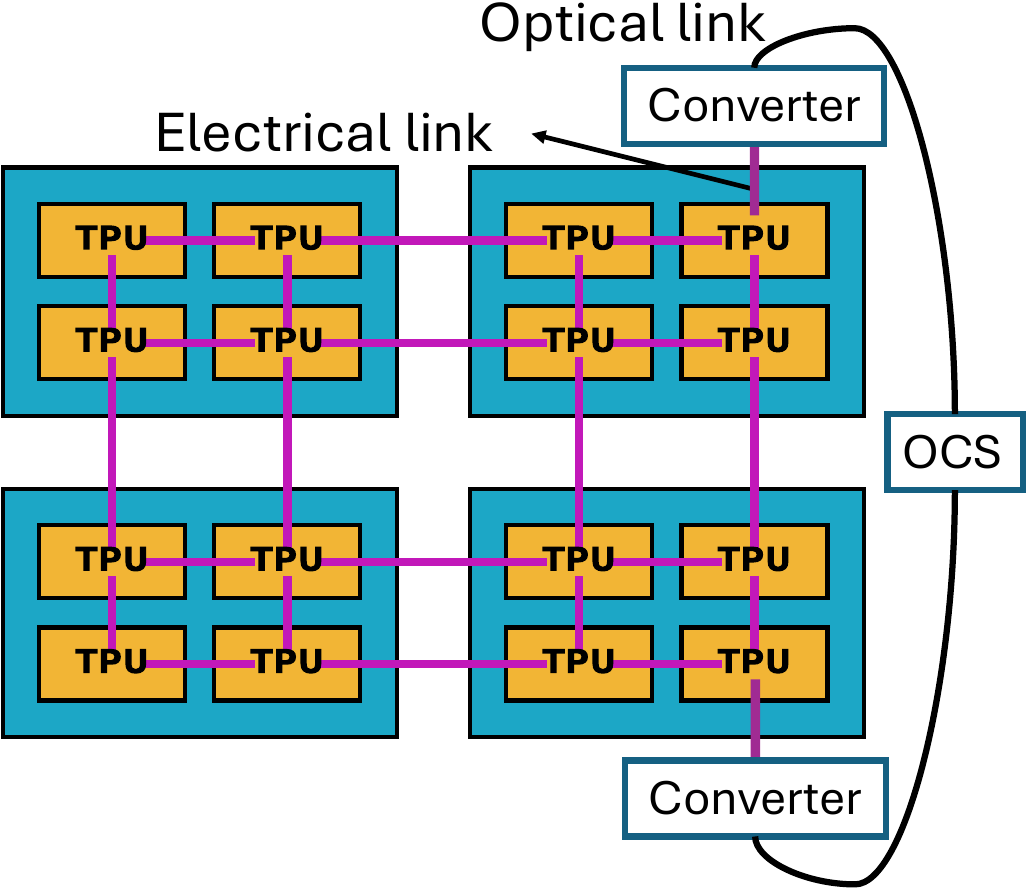}
      \caption{\small{TPU plane.}}
      \label{fig:tpu-single}
    \end{subfigure}\hfill
    \begin{subfigure}[b]{0.33\textwidth}
      \centering
      \includegraphics[width=\textwidth]{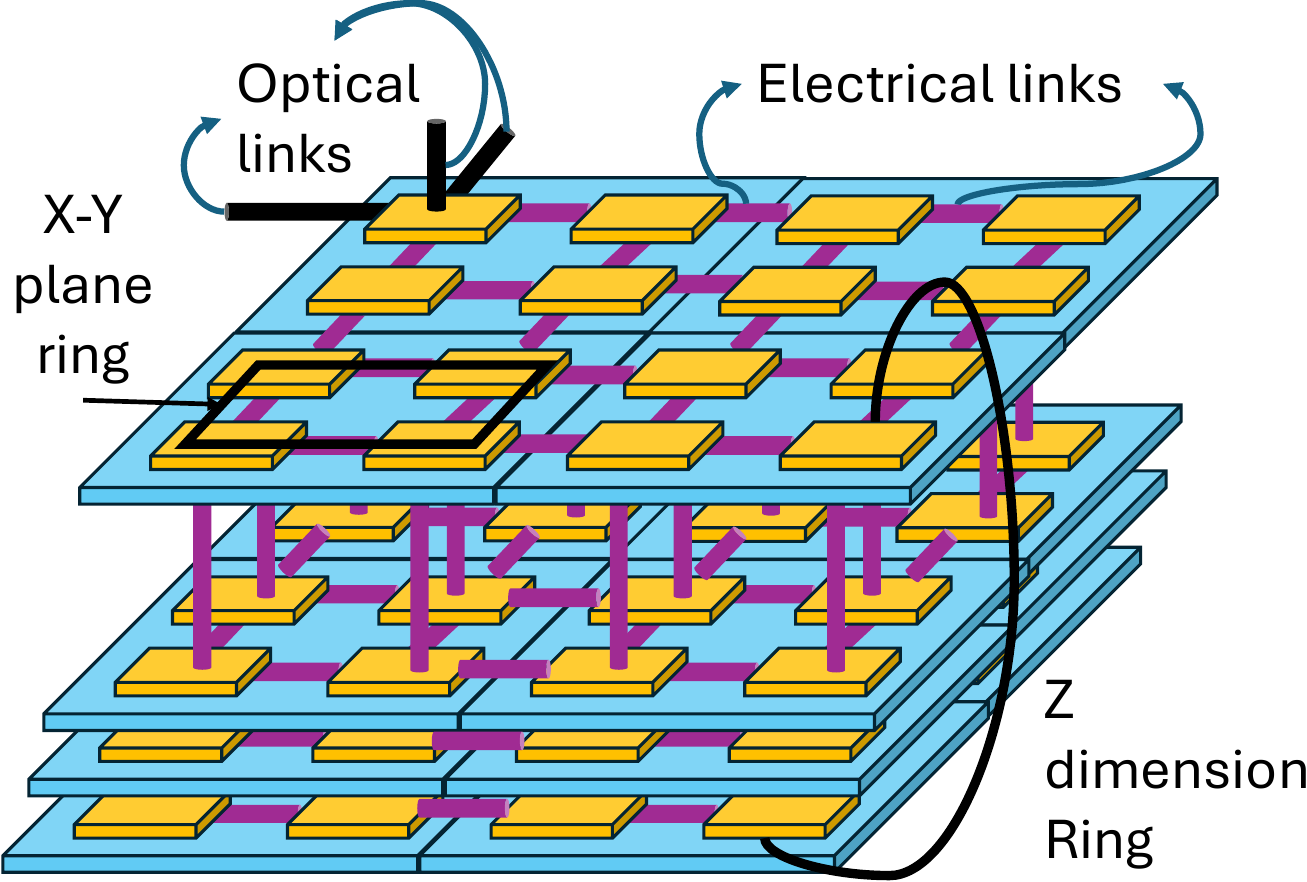}
      \caption{\small{TPU rack.}}
      \label{fig:tpu-pod}
    \end{subfigure}\hfill
    \begin{subfigure}[b]{0.32\textwidth}
      \centering
      \includegraphics[width=\textwidth]{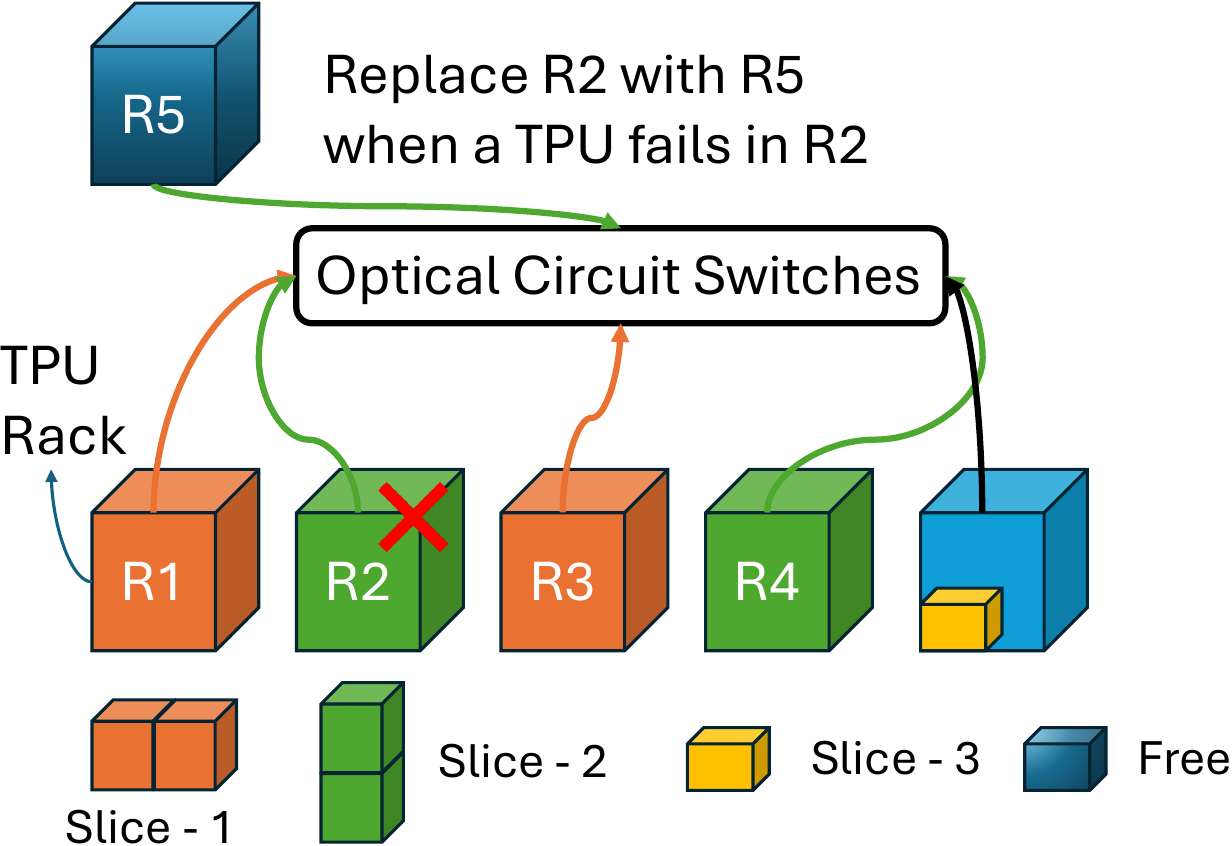}
      \caption{\small{Scale-out connections.}}
      \label{fig:tpu-slices}
    \end{subfigure}\hfill  
    \vspace{4mm}
    \caption{\small \ref{fig:tpu-single} shows a horizontal \emph{plane} of the TPU rack. This plane has 4 servers (in blue), each with 4 TPU chips (in yellow). Each TPU chip has its escape bandwidth divided across links in three directions: X, Y and Z. \ref{fig:tpu-pod} shows how multiple planes are connected to each other to form a rack. 
    Figures~\ref{fig:tpu-single},~\ref{fig:tpu-pod} show the different rings used by the collective communication algorithms optimized for the torus topology. Figure~\ref{fig:tpu-slices} shows how racks are connected to each other via OCSes to allocate multi-rack and sub-rack TPU slices to tenants.}
    \label{fig:tpu-topology}
    \vspace{-2mm} 
  \end{figure*}

In this section, we describe state-of-the-art infrastructure for large-scale ML in datacenters to put \sysname in context. 

\myparab{Multi-accelerator servers,} \eg Nvidia's DGX~\cite{dgx}, Intel's Gaudi~\cite{gaudi}, are the building blocks of large clusters and datacenter deployments used for distributed ML training and inference. A multi-accelerator server consists of a handful of ML accelerators, connected with high-speed on-board electrical interconnects (\eg ICI~\cite{googletpuv4}, Nvidia's NVLink~\cite{nvlink}, and PCIe~\cite{pcie}). Cloud operators connect racks of multi-accelerator servers into datacenter-scale deployments using a network fabric. Depending on the size of ML models, training and inference is distributed across several accelerators on the same server or across servers in a cloud datacenter.

\myparab{Limitations of traditional datacenter fabrics.} Many cloud providers use traditional datacenter fabrics to network multi-accelerator servers~\cite{vl2,jupiter-evolving,rail_optimized_2022}. In this design, optical fiber connects multi-accelerator servers in a rack to top-of-rack packet switches. The resulting racks are then connected in a Clos-like topology that is physically underpinned by optical connections but electrically packet-switched (\eg leaf-spine architecture). An ideal packet switch implements the ``big-switch'' abstraction as per which all servers connected to the switch can simultaneously communicate with each other without any resource contention~\cite{bwtax}. As the bandwidth needs of ML servers increase, it is becoming increasingly hard to build electrical packet switches that behave ideally. As a result, ML jobs in packet-switched datacenter fabrics can encounter high communication overheads due to network contention and queueing delays, reducing the throughput of ML inference and training~\cite{mlcongestion,taccl}. 

\myparab{Torus-based datacenter fabrics.} To tackle communication bottlenecks of traditional datacenter fabrics, cloud providers have deployed specialized datacenter fabrics for ML that connect multi-accelerator servers into a torus topology. Figure~\ref{fig:tpu-single} shows the horizontal cross-section of a single rack of accelerators in a torus-shaped topology~\cite{googletpuv4}. Each rack consists of 4 such planes, making each rack a torus of dimensions $4\times4\times4$ (Figure~\ref{fig:tpu-pod}). Multiple racks of multi-accelerator servers are connected to each other using optical circuit switches (OCSes) that can be programmed to establish direct connections between racks, extending the torus dimensions beyond a single rack (Figure~\ref{fig:tpu-slices}). 

\myparab{Advantages of torus fabrics.} In these datacenters, accelerators are directly connected to each other in a torus-shaped topology, avoiding packet switches. The resulting \emph{direct-connect} torus-based fabrics offer two key advantages. First, direct or point-to-point connectivity between accelerators eliminates delays due to network contention if the fabric topology matches the communication pattern of the collective communication primitive. And second, well-known collective communication algorithms offer optimal performance on certain contention-free torus topologies~\cite{karen2,googletpuv4,forestcoll,bucket-ring}. For example, the torus topology works well with multi-dimensional bucket ring algorithms~\cite{bucket-ring,googletpuv4, jain2010optimal,nd_torus}. This algorithm sequentially executes data transfers in a ring across all the dimensions of the torus ($XYZ$). Ring-based algorithms have an accelerator communicate with only two other accelerators at a given time, making communication in a ring on a direct-connect torus congestion-free.


\myparab{Example.}
Specifically, Google's torus-based datacenter consists of 64 racks, where each rack is called a \emph{pod}. Within each rack, there are 16 multi-accelerator servers, each with 4 accelerator chips. The on-board interconnect, called inter-chip interconnect (ICI), in a server is electrical. Note that \sysname replaces these electrical links within the server with programmable optical links. Bulk of the connectivity within the rack is electrical, except for \emph{wrap-around} links that connect opposite faces of the rack via optical circuit switches (green color links in Figure~\ref{fig:tpu-single}). The resulting accelerator rack forms a 3D torus~\cite{googletpuv4}. Optical circuit switches (OCSes) can be programmed to directly connect --- without packet switching --- multiple racks into larger tori (Figure~\ref{fig:tpu-pod}). 

\myparab{Tenant slices.}
A \emph{slice} is a subset of accelerators allocated to a single cloud tenant. The size of a slice ranges from a fraction of a rack ($<$ 64 accelerators) to multiple racks ($>$ 64 accelerators). Slice allocations in the torus-based datacenter must adhere to the dimensions and connectivity of a torus ($i \times j \times k$)~\cite{tpu-allocate}. Tenants deploy their training and inference jobs on the allocated accelerator slice, during which collective communication primitives are executed over the slice. However, accelerator allocations are often smaller than a single rack with 29\% of all slice allocations in the Google TPU datacenter spanning fewer than 64 accelerator chips~\cite{googletpuv4}. This highlights that there is a significant need for efficient communication on sub-rack slices.

\section{Limitations of torus-based ML fabrics}
\label{sec:motivation}

\begin{figure*}[h!]
  \begin{subfigure}[b]{0.25\textwidth}
    \includegraphics[width=\textwidth]{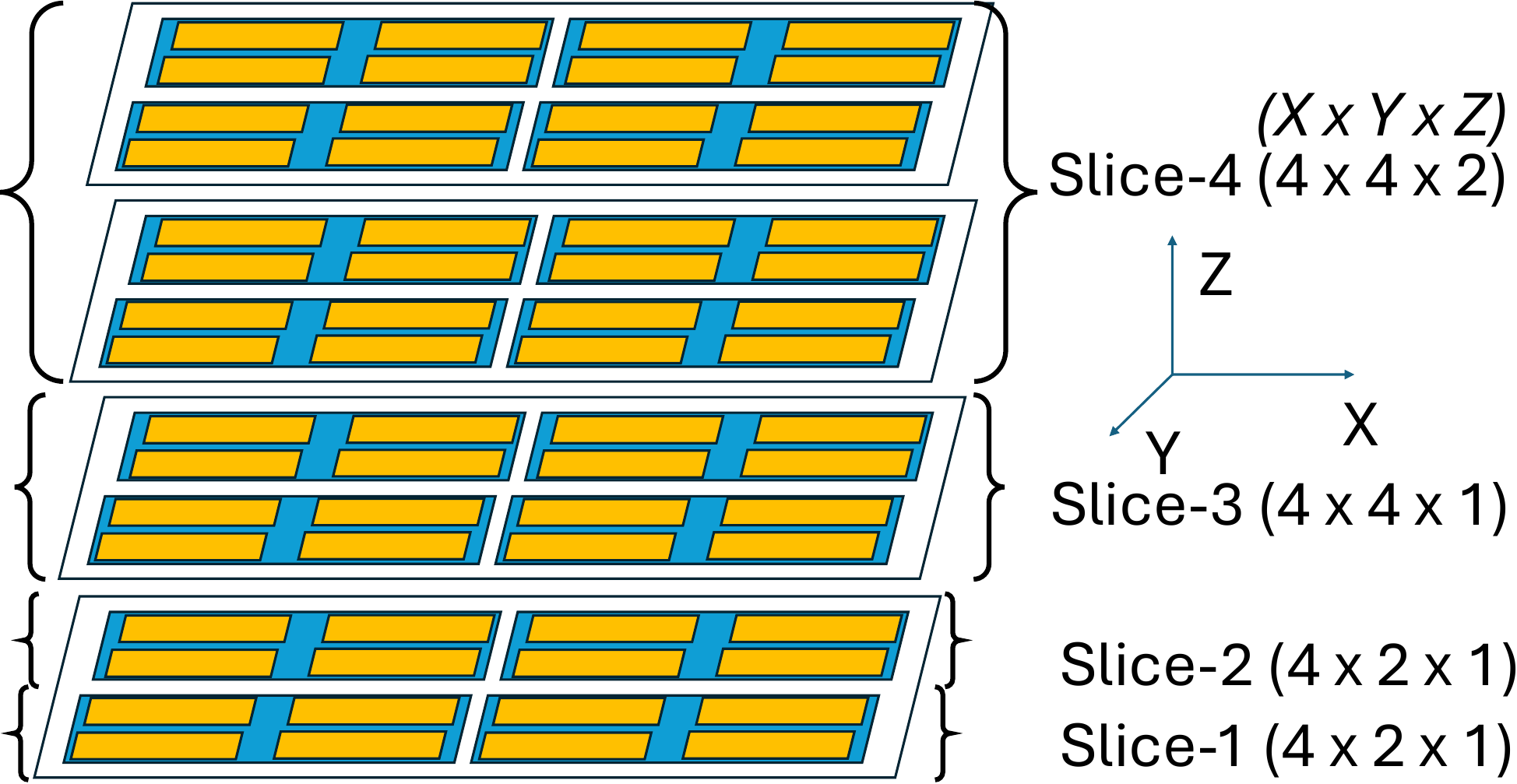}
    \caption{\small{Small slices.}}
    \label{fig:tpu_slices}
  \end{subfigure} 
  \begin{subfigure}[b]{0.22\textwidth}
    \includegraphics[width=\textwidth]{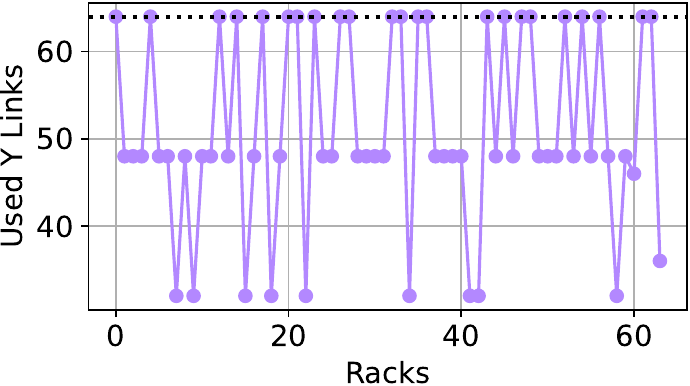}
    \caption{\small{Unused links.}}
    \label{fig:y_links_underutil}
  \end{subfigure}   
  \begin{subfigure}[b]{0.24\textwidth}
    \includegraphics[width=\textwidth]{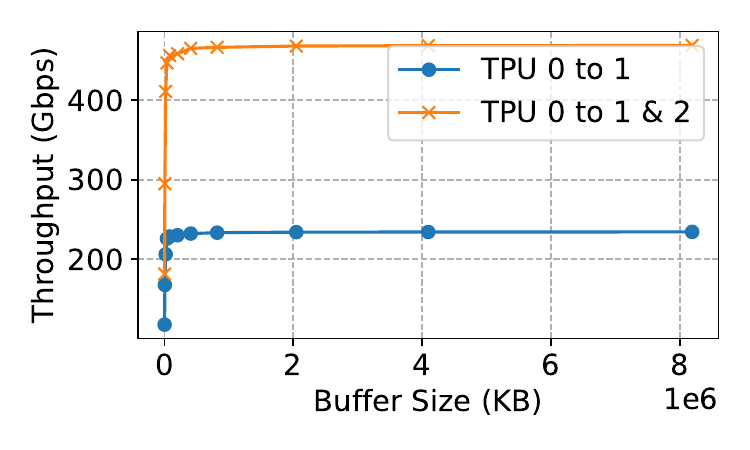}
    \caption{\small{Experimental TPU bw.}}
    \label{fig:agg_bw}
  \end{subfigure}  
  \begin{subfigure}[b]{0.24\textwidth}
    \includegraphics[width=\textwidth]{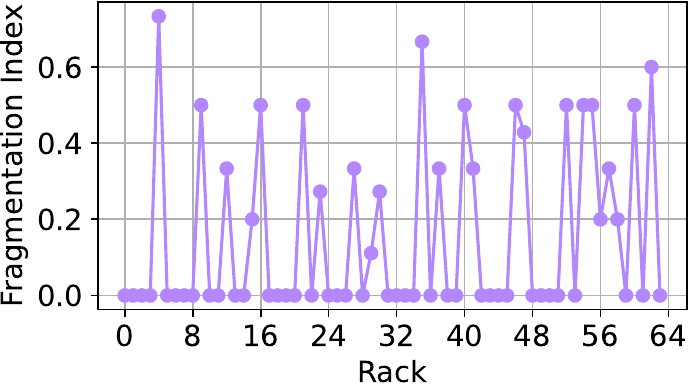}
    \caption{\small{Resource fragmentation.}}
    \label{fig:fragmentation}
  \end{subfigure}      
  \vspace{3mm}
  \caption{\ref{fig:tpu_slices} shows a rack with multiple sub-rack tenant slices.  \ref{fig:y_links_underutil} shows number of links that are used by slices in every rack (or block) in a torus-based TPU datacenter. \ref{fig:agg_bw} shows aggregate throughput of using one ICI link vs. using two ICI links. The aggregate throughput is measured through sending data in parallel to two destination devices TPU 1 and TPU 2 from TPU 0. Figure~\ref{fig:fragmentation} shows fragmentation index across racks after allocating the cluster completely and deallocating 20\% of the slices.}
  \label{fig:motivation}
\end{figure*}

In this section, we show that the performance benefits of torus-based fabrics are available only for large rack-scale compute slices~\cite{llama3_2024}. While training foundation models needs thousands of ML accelerators~\cite{llama3_2024}, inference and fine-tuning demand far fewer compute resources. Thus, it is key to understand the performance bottlenecks of \emph{any-size compute slices} in torus-based datacenters. We show, using both hardware experiments and large-scale simulations, that torus-based datacenter fabrics are susceptible to (1) significant bandwidth under-utilization (\S\ref{subsec:bw_underutil}), (2) fragmentation of compute resources (\S\ref{subsec:fragmentation_motiv}) and (3) a large blast radius of chip failures (\S\ref{subsec:fault_toler_motiv}). We will address these limitations using programmable server-scale photonic fabrics like \sysname (\S\ref{sec:design}).

\myparab{Hardware experiments and simulations.} For the hardware experiments in this section, we rent TPU slices on Google Cloud. Due to limitations on the number of TPUs we can rent (up to 32) and their versions (v3), these experiments are at a small scale, \ie up to 32 TPU chips. We develop a JAX-based tool to measure bandwidth between TPUs in the datacenter~\cite{google_jax_tpu_2024}. For the large-scale experiments in this section, we build a novel simulator of torus-based datacenter fabrics that faithfully captures the connectivity between accelerators. The simulator implements a best-effort strategy to allocate torus-shaped slices~\cite{tpu-allocate} on available chips in the simulated fabric. Our experimental tools are open-source$^{\ref{footnote:code}}$.

\subsection{L1: Bandwidth under-utilization}
\label{subsec:bw_underutil}
In a direct-connect torus topology, congestion occurs when multiple data transfers occur simultaneously on the same link~\cite{makespan1,makespan2}. Accelerators in a slice can fully use their links in a dimension only when there is no congestion in that dimension. By design, an accelerator in a 3D torus has congestion-free access to links in all three dimensions only if the slice spans an entire rack or multiple racks. Consider Figure~\ref{fig:tpu_slices} with multiple sub-rack slices. Notice the links between slices $4$ and $3$ in the Z dimension. If both tenant slices use these links, they will cause congestion on them. To avoid causing congestion, these slices must avoid using them. As a result, the bandwidth of accelerators in each slice is underutilized by 33\% because the slices have access to only 2 of the 3 dimensions. Smaller slices can have up to 66\% lower bandwidth, as we show with examples in the Appendix~\ref{appendix:examples}.

Unused inter-accelerator links in these cases are links between accelerators in \emph{different sub-rack slices}. As per the distribution of slice sizes in the TPU datacenter, sub-rack slices account for 29\% of all slice allocations~\cite{googletpuv4}. We use the distribution of slice sizes in the production TPU datacenter~\cite{googletpuv4} to allocate the entire datacenter with slices in our simulator. We find that under this representative load, a significant number of links remain unused. Figure~\ref{fig:y_links_underutil} shows severe link underutilization, \ie up to 50\% of the links in the Y dimension are unused in some racks of the cluster. 

\myparab{Low slice bandwidth.}
The challenge of unused links is two-fold. First, significant amount of provisioned bandwidth is underutilized as shown in Figure~\ref{fig:y_links_underutil}. Second, accelerators in sub-rack slices can use only a fraction of the bandwidth they physically support since the egress bandwidth of the accelerator is statically partitioned across three dimensions. If the slice shape prevents the accelerator from using a certain dimension, bandwidth in that dimension remains unused. 

\myparab{Redirecting bandwidth based on slice shape.} 
\sysname implements the hardware and software capability (\S\ref{sec:design}) to redirect idle bandwidth of accelerators where it is needed. In case of idle bandwidth due to the shape of the tenant slice, we will redirect bandwidth in unused torus dimensions along connections in the tenant slice. We experimentally verify that using links in more dimensions improves the egress bandwidth accelerators in the TPU datacenter. We rent TPU slices in Google cloud and use \texttt{Jax.device\_put} to measure the throughput of sending and receiving data among TPU accelerators in a $2\times2\times1$ TPUv4 slice. The aggregate throughput is measured through sending data in parallel to two adjacent devices of TPU 0, TPU 1 (along x axis) and TPU 2 (along y axis) from TPU 0. Our experiment shows that the aggregate throughput of two links is $2\times$ the throughput of one link for a TPUv4 accelerator (Figure \ref{fig:agg_bw}). Thus, a slice that consists of only TPU 0 and 1 could have 2$\times$ bandwidth by using the redirected cross-slice bandwidth between TPU 0 and 2 in a 2D torus. This also shows that the I/O bandwidth of the accelerator is not the limiting factor since the network interface supports concurrent communication across all dimensions.


\subsection{L2: Compute fragmentation}
\label{subsec:fragmentation_motiv}
In traditional packet-switched datacenter fabrics~\cite{vl2,rail_optimized_2022}, allocating a group of multi-accelerator servers anywhere in the datacenter to a single tenant is trivial, since each server connects to the datacenter fabric through its own top-of-rack switch. Thus, while susceptible to congestion delays, packet-switched fabrics allow flexible allocation of compute to tenants. Contrast this with direct-connect torus fabrics~\cite{googletpuv4}. To allocate XPU servers to a tenant, the cloud provider must find a \emph{contiguous} set of unused XPUs whose connectivity matches that of a torus of desired dimension~\cite{tpu-allocate}. This constraint arises because allocating a tenant slice using non-contiguous XPUs necessitates intra-slice communication to route through other tenant slices, causing congestion in direct-connect deployments. Thus, \emph{compute fragmentation} occurs when unused servers in the datacenter cannot be allocated due to non-contiguity of their location.

Specifically, allocations and deallocations of sub-block slices across the photonic datacenter fabric over time leads to fragmentation of resources within a rack. We quantify the fragmentation index $I$ of a rack as: $I = 1 - \frac{S}{T}$, where $S$ is the number of XPUs in the largest allocatable slice, and $T$ denotes the total number of unallocated XPUs within the rack~\cite{silberschatz2018operating}. In our simulator, we fully allocate the XPUs in a cluster by sampling sub-block slice sizes from the publicly available production distribution~\cite{googletpuv4} and then deallocate 20\% of the slices at random. Figure~\ref{fig:fragmentation} shows fragmentation index across all 64 racks in a XPU cluster in our experiment. In the worst case, racks are susceptible to high fragmentation, as high as $I=0.7$, which hampers the ability to allocate slices to tenants, despite the availability of free accelerators.

\subsection{L3: High blast radius of accelerator failures}
\label{subsec:fault_toler_motiv}

Chip failures are common in ML-centric datacenter deployments~\cite{megascale}. In fact, Meta reported that 30\% of the failures during the training of Llama 3.1 were due to faulty GPUs~\cite{llama3_2024}. Naturally, cloud operators employ several fault tolerance policies in ML clusters. For instance, Google's TPU supercomputer migrates a multi-rack ML job away from the rack with a failed TPU chip to a different rack. In addition to migrating the job, the new set of TPU racks is directly connected --- without inter-rack electrical packet switching --- by reconfiguring the optical circuit switches that link all TPU racks~\cite{googletpuv4,tpu-resilience}. Even so, reconfigurable datacenter fabrics have an excessively large blast radius, \ie the extent of impact of the failure of a single accelerator chip. Not only is the migration expensive for the job interrupted by failure, but it may also be infeasible to find an entirely unused set of servers for every job affected by a single failed XPU.

\section{Server-scale programmable fabrics}
\label{sec:hardwaredesc}

We address limitations (L1---L3) of torus-based datacenters (\S\ref{sec:motivation}) by developing a programmable photonic chip-to-chip fabric or interposer, \sysname. In \sysname, accelerators are stacked on top of the server-scale photonic fabric (Fig.~\ref{fig:wafer}). The egress bandwidth of the accelerator via SerDes ports feeds into a \emph{silicon photonic mesh} of waveguides and optical switches~\cite{lumorphofc}. The key advantage of programmable chip-to-chip optical fabrics is their ability to dynamically redirect egress bandwidth of an accelerator along connections where it is needed instead of statically partitioning it among a subset of inter-GPU links, as done in direct-connect tori. 

We omit a detailed discussion of silicon photonic interposers due to limited space but briefly describe how \sysname interacts with the mesh. At a high level the programmable optical interposer brings the ability to ``switch'' bandwidth near the compute unit \ie XPU, right after the output of the SerDes reaches the optical Tx/Rx of the mesh (Fig.~\ref{fig:wafer}). By programming switches on the mesh, \sysname redirects bandwidth from an accelerator to any other accelerator in the same rack. This bandwidth traverses on optical circuits between accelerators which maintain direct-connectivity since there are no packet switches on the all-optical path between accelerators. Creating circuits involves finding routes on the silicon photonic mesh and we leverage related work to find viable routes~\cite{ding2025pipswitchcircuitswitchusing}. In this manner, \sysname tackles limitations of direct-connect fabrics by:

\setlist{nolistsep}
\begin{compactitem}
  
  \item (L1) Eliminating bandwidth underutilization in sub-rack slices by redirecting idle bandwidth to connections within the tenant slice, allowing accelerators to use full escape bandwidth regardless of slice shape and size.
  \item (L2) Eliminating compute fragmentation by creating intra-rack circuits between physically non-contiguous accelerators, making them logically contiguous without incurring contention on links in other slices.
  \item (L3) Minimizing blast radius of accelerator failures by redirecting bandwidth from accelerators neighboring the failed one towards a healthy accelerator in the same rack.   
\end{compactitem}

\myparab{Can packet switches solve these challenges?}
One might argue that connecting accelerators within the server using an \emph{ideal packet switch} can achieve the same effect: contention-free connections between all accelerators~\cite{bwtax,bigswitch}. However, inter-accelerator bandwidth within modern servers is already massive --- over 900 gigabytes per second in one direction~\cite{nvlink}--- making it harder to stay true to the ideal switch abstraction. This has already resulted in evidence of contention in switched server-scale interconnects~\cite{hostcongestion,taccl}.

\myparab{Programming circuits infrequently.}
Creating optical circuits by programming silicon photonic switches incurs a delay of several microseconds~\cite{hotchips34}. Therefore, \sysname creates new circuits only in two cases (1) at the time of allocating a new slice to tenants and (2) to handle sudden accelerator failures. Both scenarios are tolerant to the microsecond-scale reconfiguration delay of server-scale photonic fabrics.

\section{\sysname's software orchestrator}
\label{sec:design}

We develop a software orchestrator, \orchestrator, for direct-connect ML datacenters augmented with server-scale photonic fabrics like \sysname.
\orchestrator has 3 components:

\setlist{nolistsep}
\begin{compactitem}

  \item \allocator : given a tenant's request for a slice of compute resources in the datacenter fabric, \allocator finds an unallocated set of physical resources that match the slice dimensions and allocates them to the tenant. The \allocator keeps state of available datacenter resources, OCS configurations and server-scale fabric configurations. It allocates compute such that accelerators in the tenant slice have access to their full egress bandwidth, addressing the limitation of idle bandwidth discussed in \S\ref{subsec:bw_underutil}. The \allocator logically connects fragmented resources to allocate slices if contiguous resources are not available, addressing the fragmentation limitation discussed in \S\ref{subsec:fragmentation_motiv}.
  \item \faultmanager : \faultmanager reacts to an accelerator failure by identifying an unallocated healthy chip within the same rack and logically replacing the failed chip, addressing the limitation in \S\ref{subsec:fault_toler_motiv}. This \emph{in-place} replacement of faulty chips is possible by connecting the healthy chip via optical circuits to the remaining slice.
  \item \controlplane : Both \allocator and \faultmanager identify the logical connectivity needed for a tenant slice. Their output is fed to the \controlplane. The \controlplane translates logical slice configurations into physical configurations of the \sysname silicon photonic meshes in the datacenter.
\end{compactitem}

\begin{figure}[h!]
    \begin{center}
        \includegraphics[width=0.45\textwidth]{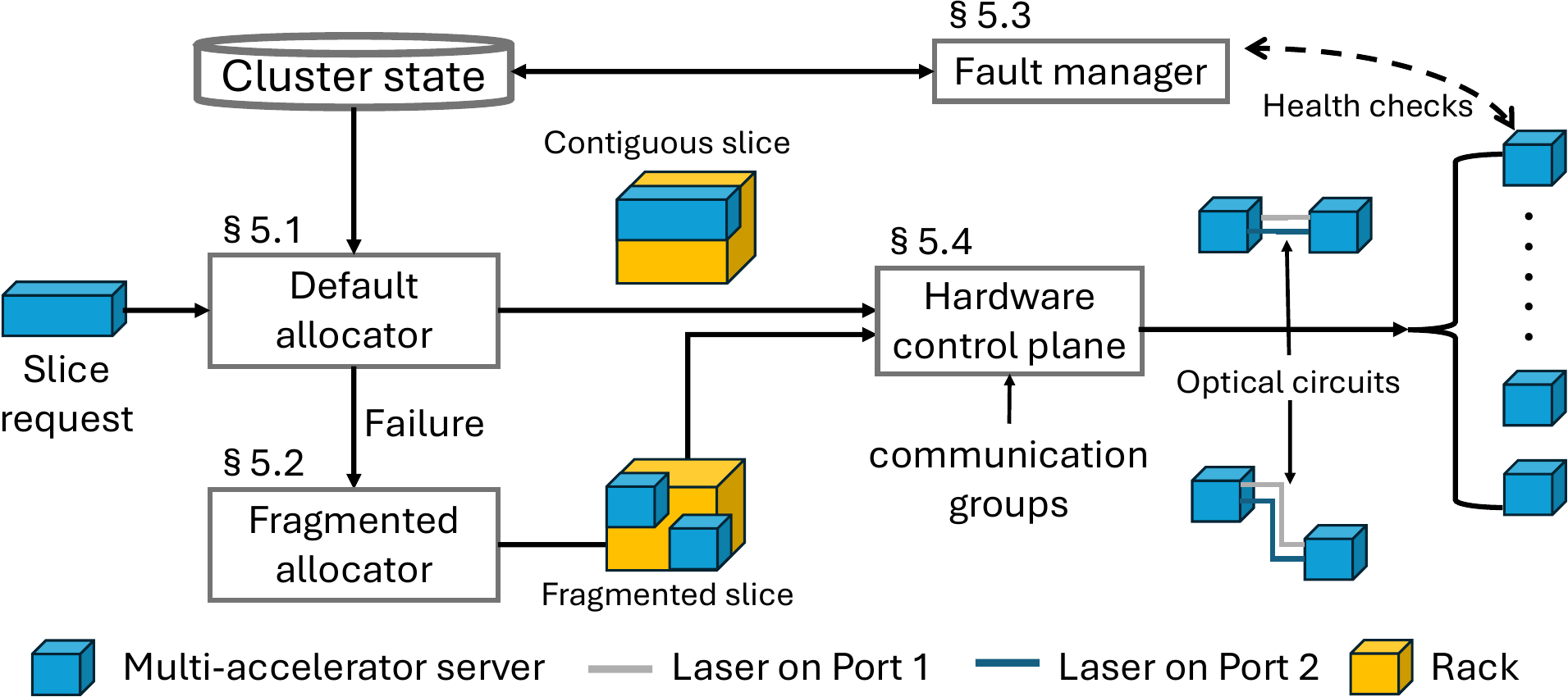}
    \end{center}
    \vspace{-1mm}
    \caption{Design of \orchestrator.}
    \label{fig:mgr} 
  \end{figure}

Figure~\ref{fig:mgr} summarizes the 3 components of \orchestrator. Next, we describe resource allocation algorithms implemented by the \allocator, \faultmanager and \controlplane. 

\subsection{Resource allocator}

Fig.~\ref{fig:mgr} shows the architecture of \sysname's resource allocator. Input to the resource allocator is a slice request represented using the tuple $x \times y \times z$. The goal of the resource allocator is to find enough XPUs in the cluster and connect them in the slice's torus  topology. The output of the allocator is the coordinates of each XPU in the rack and the links connecting XPUs. Each link is represented with a list of coordinates from source to the destination including hops, if any. The allocator tries to satisfy slice requests with contiguous XPUs in the rack that are connected in the same topology as the slice request. For example, a $2\times2\times2$ slice will have a 1:1 mapping to a $2\times2\times2$ portion of the 64 XPU 3D torus rack. If no such physically contiguous servers exist, the allocator uses the fragmented allocator described next.


\subsection{Fragmented slice allocator}
\label{sec:fragment_alloc}
Fragmentation of resources necessitates connecting TPUs in non-contiguous servers of a rack with optical circuits to create the desired slice topology. To do this, the \allocator creates circuits between \sysname servers over optical fibers. We formulate the problem of allocating slices from non-contiguous servers as an integer linear program (ILP) which finds the optimal placement for slices while minimizing the number of fiber paths needed between servers.

\myparab{Slice request topology.} 
Users request slices in a torus topology with x,y and z TPUs in each of the respective axes denoted by the graph $G: \langle L, T \rangle$. $L$ is the set of servers in the slice request which we refer to as slots and $T$ is the edges connecting these slots forming the slice topology. 

\myparab{Hardware topology.} We denote the rack with graph $G': \langle S,I \rangle$. $S$ is the set of physical servers in a rack and $I$ are the optical fibers connecting consecutive servers. ${P}(u, v)$ is the set of pre-computed paths between physical servers $u$ and $v$, and can be sampled with a variety of algorithms~\cite{yens-algo}.

\myparab{Decision variables.}
Algorithm~\ref{alg:fragmented_allocator} uses $x_{a,b}$, a binary variable to indicate if server $b \in {S}$ is mapped to slot $a \in {L}$ and
uses binary variable $r_{i,u,v}$ to indicate if path $i \in {P}(u, v)$ is selected for communication between servers $u$ and $v$.

\myparab{Why allocate at server granularity?} Server instead of TPU granularity in decision making reduces the computation overheads without loss in solution quality. Our key insight is that the number of fibers attached to a server is much fewer than the number of waveguides in a \sysname server and thus we can route circuits within the \sysname fabric as long as the inter-fabric links can accommodate the circuits.


\myparab{Mapping constraints.}
We only allocate one server per slot from the set of free servers $F \subset S$ with these constraints: (1) $\sum_{a \in {L}} x_{a,b} \leq 1, \text{ } \forall b \in {F}$; (2) $\sum_{a \in L} \sum_{b \in F} x_{a,b} == len(L)$.



\myparab{Path selection.}
We find paths between servers if they are assigned to slots in the slices that are connected within the slice. For all $m, n \in F$ and $a, b \in L$, if $x_{a,m} \cdot x_{b,n} = 1$ and edge $e_{a,b} \in T$, then $\sum_{i \in P(m, n)} r_{i,m,n} = 1$.



\myparab{Objective.} The algorithm minimizes $z$, which is the maximum number of circuits passing between two servers. Every pair of servers directly connected in any dimension have 4 fibers between them (Figure~\ref{fig:tpu-single}). We assume every optical circuit uses all wavelengths to model the worst case fiber need. So, for all edges in the rack, $\forall{e \in I}$ and for all routes $Q$ passing through this edge $Q \in P$, $z \geq \sum_{i \in Q}$ $r_{i,u,v} \cdot 4 + b(e)$, where $b(e)$ is the number of already existing circuits on $e$ from other slices. Algorithm~\ref{alg:fragmented_allocator} in the Appendix summarizes the ILP formulation. We show in the evaluation (\S\ref{sec:sim}) that the ILP can be solved in a milliseconds for typical slice requests.


\subsection{In-place fault tolerance with \sysname} 
\label{subsec:inplace_ft}

\begin{figure*}[t] 
  \centering
    \begin{subfigure}[b]{0.25\textwidth}
      \includegraphics[width=\textwidth]{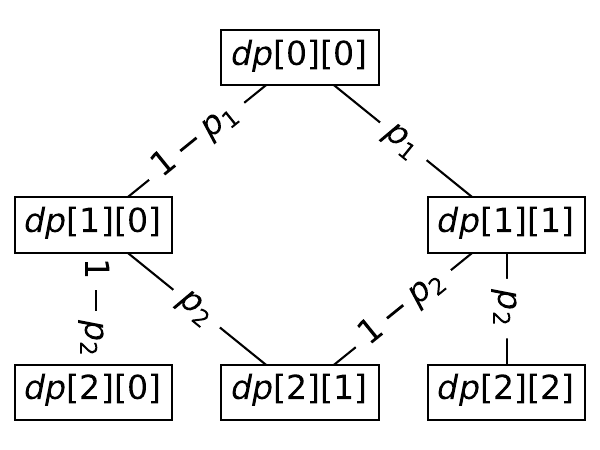}
      \caption{\centering \small Tree of $dp$ matrix.}
      \label{fig:dp}
  \end{subfigure}\hfill
  \begin{subfigure}[b]{0.3\textwidth}
    \includegraphics[width=\textwidth]{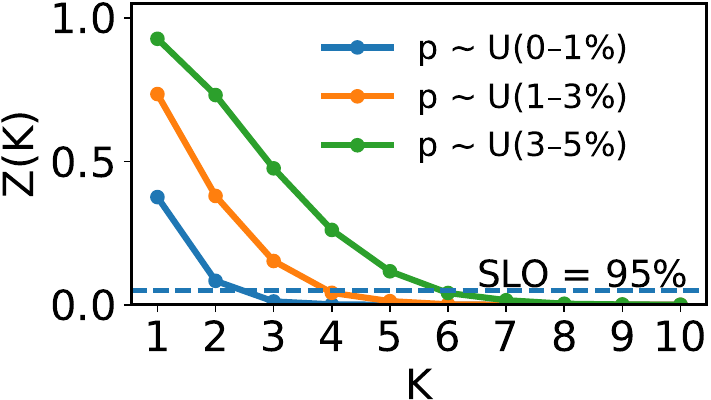}
    \caption{\centering \small SRG = XPU}
    \label{fig:zk-failures}
\end{subfigure}\hfill
\begin{subfigure}[b]{0.3\textwidth}
  \includegraphics[width=\textwidth]{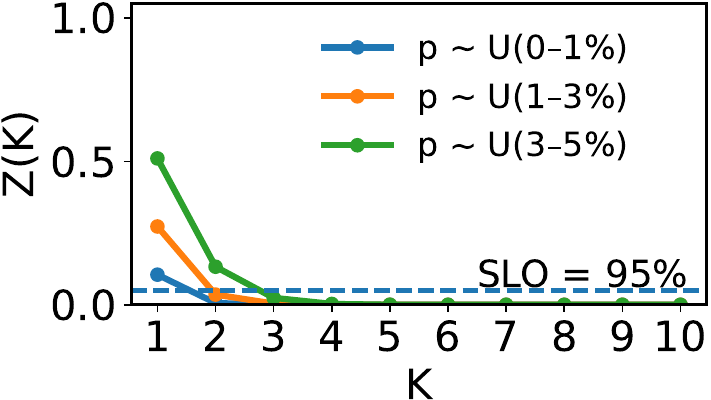}
  \caption{\centering \small SRG = Server with 4 XPUs}
  \label{fig:zk-failures-servers}
\end{subfigure}\hfill
\label{fig:failures}
\vspace{3mm}
\caption{Figure~\ref{fig:dp} shows the recursion tree of Z(K) for 2 SRGs. In Figure~\ref{fig:zk-failures}, we set SRG to a single XPU and N to 64. shows that with 4 XPUs are sufficient to respect $95\%$ SLO when SRG is a single XPU. Similarly Figure~\ref{fig:zk-failures-servers} shows that two additional servers (4 XPUs per server) are sufficient to respect $95\%$ SLO in most cases. In both, ~\ref{fig:zk-failures} and~\ref{fig:zk-failures-servers}, need for additional hardware increases with failure probability.}
\end{figure*}

We design a \faultmanager in \sysname to tolerate faults without hardware overprovisioning, as done in torus-based fabrics, or inducing network congestion. To achieve this, we must answer the following research questions:

\begin{compactitem}
\item \myparab{How many XPUs to reserve for fault-tolerance?} To prevent overprovisioning, it is essential to determine the number of additional XPUs required for fault tolerance. The number of XPUs to add is lower bounded by the number of XPUs that can simultaneously fail. We find the optimal number of XPUs by determining the highest number of XPUs that can fail within a given Service Level Objective (SLO). \eg for an SLO of 95\%, we need enough free XPUs to tolerate failure scenarios with a probability > 5\%.

\item \myparab{Where should we place the spare XPUs?} New optical connections made to spare XPUs after a failure must be routed over distinct fibers to preserve signal integrity and avoid congestion, often requiring additional fibers between servers in the rack. Therefore, it is crucial to place spare XPUs such that it reduces the number of new fibers needed in each rack. We add extra XPU servers instead of reserving ones in the torus rack to preserve the rack's original dimensions. Reserving a server anywhere within a rack, reduces the effective size of the rack (\eg from $4 \times 4 \times 4$ to $4 \times 4 \times 3$ if we reserve a corner server), which breaks support for multi-rack slices with sizes that are multiples of $(4 \times 4 \times 4)$. 
\end{compactitem}


\myparab{Shared risk groups (SRG).}
Components in a datacenter follow the fate-sharing principle~\cite{fate_sharing} which suggests that related parts of a system fail together or none of them fail. We use the popular concept of a Shared Risk Group (SRG), a group of resources that fail together. Each SRG fails independently of others since an SRG consists of all resources that have shared dependencies (aside from datacenter-wide failures). For instance, if an XPU's compute cores or memory fail, the entire XPU is treated as failed. Similarly, a failure in a server's power supply, host CPU, or DRAM results in the failure of the entire server, including all the XPUs it hosts. 

\myparab{Failure probability of a SRG.}
We define the probability of failure of an SRG as $P_{\text{fail}} = \frac{T_{\text{repair}}}{T_{\text{active}} + T_{\text{repair}}}$ where $T_{\text{repair}}$
is the failure duration of the SRG, including the time taken to repair it, $T_{\text{active}}$ is the duration of time when the SRG is healthy. This duration should be extracted from the logging, telemetry and diagnostic collection pipeline of the cloud provider.

\myparab{Probability $\geq K$ failures.}
We define $Z(K)$ as the probability that 
$\geq K$ SRGs fail in $N$ SRGs. Let $p_i$ be the $P_{\text{fail}}$ of the $i^{th}$ SRG, then $Z(K) = \sum_{i=K}^{N} \sum_{\substack{A \subseteq \{1, \ldots, N\} \\ |A| = i}} \left[ \prod_{j \in A} p_j \right] \left[ \prod_{j \notin A} (1 - p_j) \right]$. $Z(K)$ is expensive to compute as it iterates over all subsets of sizes $K$ to $N$. 

    

\myparab{Key insight:} Our key insight is that the inner sum of $Z(K)$ can be represented as a recursive relation with smaller sub-problems. Let $dp[i][k]$ be the probability of $k$ failures in $i$ SRGs, then the recursive relation is given by $dp[i][k] = dp[i-1][k-1] * (p_{i}) + dp[i - 1][k] * (1 - p_{i})$. We formulate this as a dynamic program which computes larger $dps$ using smaller solutions as shown in Figure~\ref{fig:dp}. Given the $dp$ matrix, we can compute Z(K) by summing over all probabilities where at least $K$ failures occur, i.e., $Z(K) = \sum_{k=K}^{N} dp[N][k]$. This reduces the computational complexity from $O(2^N)$ to polynomial $O(N^2)$, making it tractable for large networks with many SRGs. We pick 
$K$ SRGs such that $Z(K) \leq (1 - S)$ where $S$ is $\frac{SLO}{100}$. This $K$ guarantees that the probability of $K$ or more SRGs failing is less than $(1 - S)$. So, all the failures
that occur during the remaining $S$ fraction of the time can be gracefully handled by adding $K$ more SRGs.


\myparab{How do we pick N?} We calculate $Z(K)$ at the rack granularity because adding more XPUs not only requires adding more compute but also networking resources to connect these XPUs to the existing XPUs. One can connect additional XPUs to the OCSes and replace failed XPUs in any rack by the additional XPUs. However, the ports on the OCSes connecting multiple racks are already saturated which makes connecting additional XPUs to the OCSes infeasible. Cascading multiple OCSes to increase the switch radix is also challenging~\cite{tpu-resilience} due to optical losses. In Figure~\ref{fig:zk-failures}, we set $N=64$ (the number of XPUs in a rack), SRG to a single XPU and sample the failure probabilities of each XPU from three different ranges. The plot shows that reserving just 4 XPUs handles more than 95\% of the failures in most cases. We also set SRG to an entire server with 4 XPUs and calculate $Z(K)$ in Figure~\ref{fig:zk-failures-servers} which shows that 2 servers are sufficient to handle most failures in each rack. 

\myparab{Placing the additional XPUs.}
We analyze the placement of additional XPUs by setting SRG to a single XPU. Given that a server contains four XPUs—sufficient for handling over 95\% of failures, we optimize placement at server granularity. Each rack consists of 16 servers arranged in a torus. Leveraging symmetry, the unique positions where a new server can be added are $(-1,0,0),(0,-1,0),(0,0,-1),(0,-1,1),(-1,0,1)$.

\myparab{Fault detection and response.}
The \faultmanager communicates 
with daemons running on every multi-accelerator server which report the status of processes running on the XPUs. Upon detecting failure, \faultmanager searches for a free and healthy XPU and invokes \controlplane to create optical connections between the neighbors of the failed XPU and the healthy XPU to logically replace the failed XPU. Finally, it restarts the job with the latest checkpoint~\cite{checkfreq}. We provide implementation details of both \faultmanager and \controlplane in Appendix \ref{appendix:implementation}.

\subsection{Hardware control plane}
The \allocator and \faultmanager output a set of servers and circuits between them that form the desired slice topology. These optical circuits connect XPUs that are involved in collective communication resulting from various parallelisms. For example, a XPU might be a part of two rings -- one ring for Tensor Parallelism (TP) and one ring for Data Parallelism (DP) which requires four optical circuits, two for each ring. The \controlplane is responsible for implementing these optical circuits on the underlying optical fabric using route-finding algorithms which we implement using the approach from related work~\cite{ding2025pipswitchcircuitswitchusing}.
\begin{figure*}[h]
    \centering 
    \begin{subfigure}[b]{0.4\textwidth}
    \includegraphics[width=\textwidth]{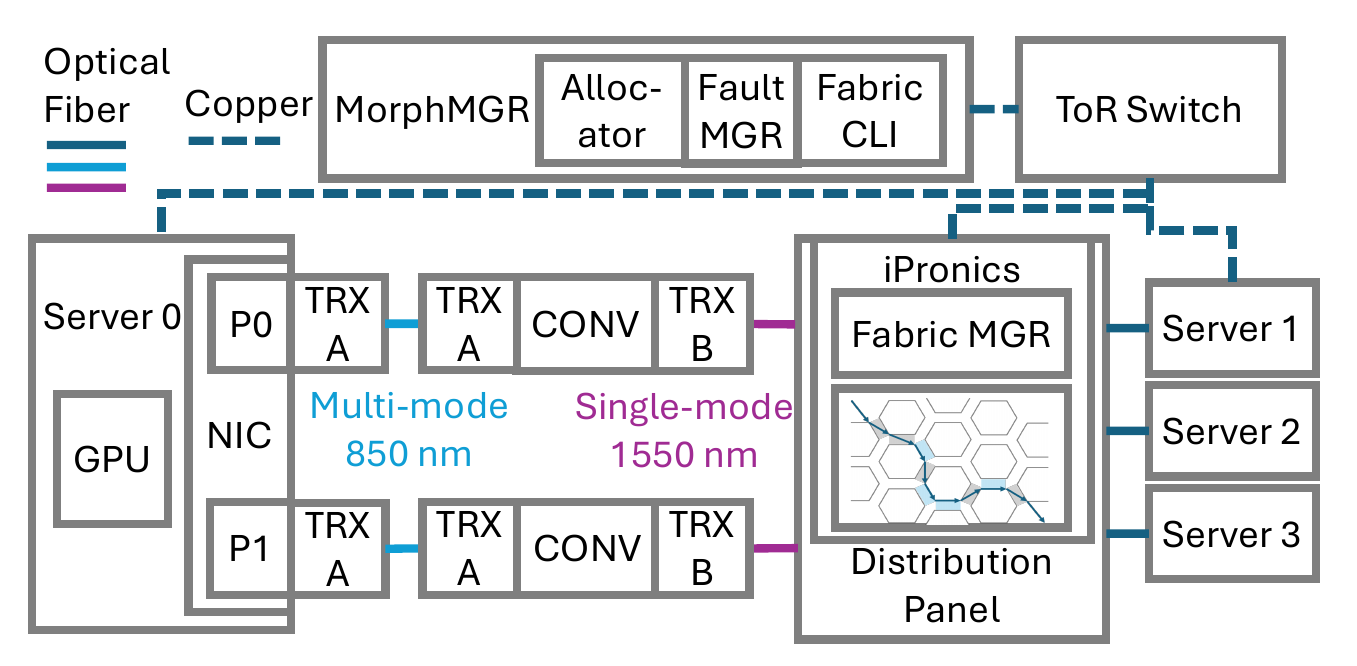}
    \caption{\small{Diagram of the testbed.}}
      \label{fig:hardware}
    \end{subfigure} 
    \begin{subfigure}[b]{0.24\textwidth}
        \includegraphics[width=\textwidth]{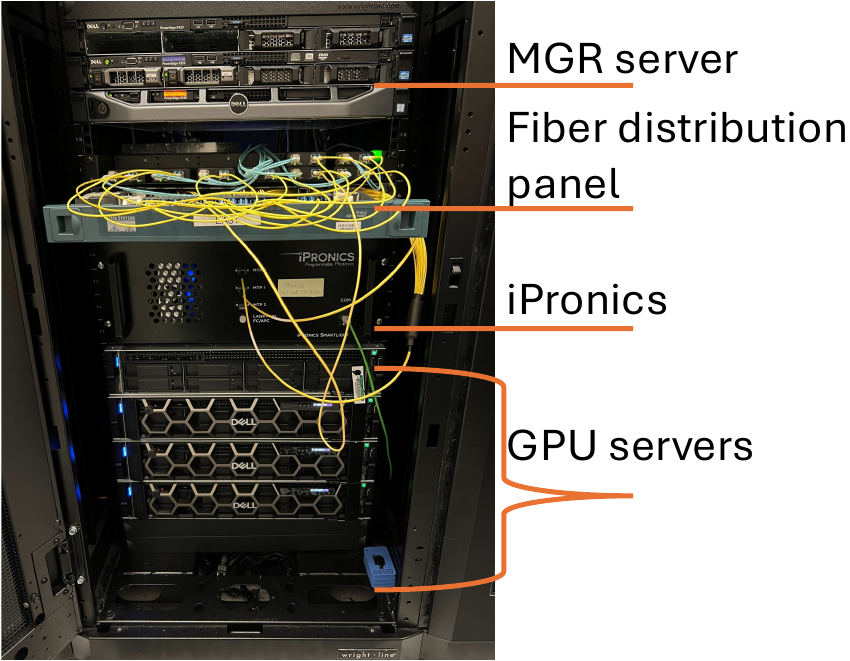}
        \caption{\small{Photograph of the testbed.}}
      \label{fig:hardware_testbed}
    \end{subfigure}  
    \begin{subfigure}[b]{0.3\textwidth}
        \includegraphics[width=\textwidth]{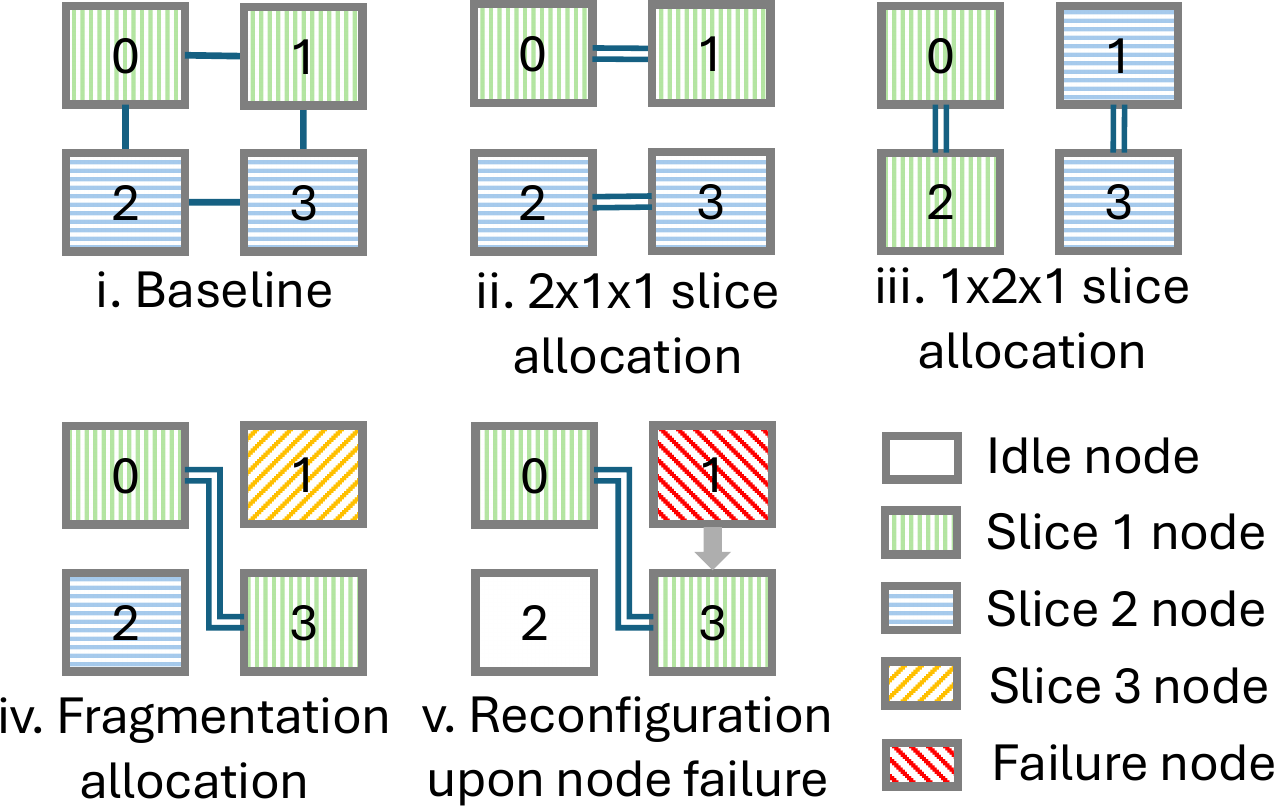}
        \caption{\small{Topologies evaluated.}}
      \label{fig:hardware_topo}
    \end{subfigure} 
    \vspace{3mm}
    \caption{\ref{fig:hardware} shows the diagram of \sysname prototype (TRX: stands for FS 10GBASE-SR SFP+ 850nm transceivers \cite{fs_tr_850}; TRX B: FS 10GBASE-ER SFP+ 1550nm transceivers \cite{fs_tr_1550}; CONV: FS 1x10G SFP+ to 1x10G SFP+ fiber media converters \cite{fs_conv}). \ref{fig:hardware_testbed} is a photo of the testbed. And \ref{fig:hardware_topo} shows a list of topologies being used in the evaluation.} 
     \label{fig:hardwareeval}
  \end{figure*}

\section{Hardware evaluation of \sysname}
\label{sec:hardwareeval}

We build an end-to-end hardware prototype to evaluate \sysname. Adding accelerators to the \sysname wafer-scale interconnect (\S\ref{sec:hardwaredesc}) would need using a semiconductor fabrication facility like TSMC~\cite{tsmc} to stack chips on the interconnect. Semiconductor packaging at the small scale of one prototype like this incurs the cost of several million dollars, which is prohibitively expensive for academic research. We overcome this challenge using an off-the-shelf silicon photonic mesh~\cite{lopez2019programmable,perez2018programmable}, \ipronics, that uses the same switch and waveguide technology as used in optical interposers (Figure~\ref{fig:wafer}). Instead of using advanced packaging, this mesh connects to XPUs directly by exposing waveguides via fiber connectors which can communicate via optical transponders.

We connect four GPU servers to \ipronics via their NICs and optical transceivers. While the datapath on \ipronics is GPU $\rightarrow$ PCIe $\rightarrow$ NIC $\rightarrow$ photonics compared to GPU $\rightarrow$ photonics on \passage in Figure~\ref{fig:wafer}, \ipronics offers the same programmability as \passage once the data reaches it from the GPU. We found an undocumented limitation of the \ipronics mesh  --- it only accepts light with transverse-electric polarization. In contrast, all commodity transponders emit polarization multiplexed light. Due to this mismatch, connections between GPUs and the mesh were highly lossy. Despite the capability of the mesh to carry terabits of traffic, we could only test it with 10 Gbps connections~\cite{ding2025pipswitchcircuitswitchusing}. 

\myparab{Testbed setup.}
The testbed shown in Figures \ref{fig:hardware} and \ref{fig:hardware_testbed} comprises four Dell 7960 rack servers running Ubuntu 24.04.1 LTS. Each server features an NVIDIA RTX 6000 Ada GPU with 48GB of memory \cite{NVIDIA_A6000}, a Broadcom BCM5720 NIC with two 1GbE ports, and an Intel E810-XXVDA4 NIC with four 25/10/1 GbE ports \cite{intel_nic}. One Broadcom port connects to a top-of-rack switch for management and control. Two Intel ports connect to a programmable silicon photonic fabric with a mesh of hexagonal waveguides (\ipronics) \cite{lopez2019programmable}. For interoperability, we use FS 10GBASE-SR SFP+ 850nm transceivers \cite{fs_tr_850} on the intel NICs and convert it to 1550nm using FS 1x10G SFP+ to 1x10G SFP+ fiber media converters \cite{fs_conv} and FS 10GBASE-ER SFP+ 1550nm transceivers \cite{fs_tr_1550} and feed this 1550nm signal to \ipronics via optical fibers.

\myparab{Baseline.}
We build a 2D torus with the 4 servers in our testbed, shown in Figure~\ref{fig:hardware_topo}(i) to reflect the torus topology adopted by the TPU cluster. We statically partition the two NIC ports across the X and Y dimensions in the baseline and perform no reconfigurations to model the default behavior of the electrically interconnected TPU rack. We also use the TPU's migration policy to tolerate faults in the baseline.



\myparab{Workloads.} We use \texttt{iperf} and \texttt{nccl\_test} to evaluate the available bandwidth. To demonstrate end-to-end benefits, we fine-tune \texttt{Llama-3.2-1B} \cite{llama3.2} on the wikitext~\cite{wikitext} dataset, using the maximum batch size the GPU's memory permits and distributed data parallelism across 2 GPUs. We use shapes $2\times1\times1$ and $1\times2\times1$ in slice requests.  Every slice contains one communication group, and all available bandwidth is reserved for this group. The \controlplane redirects bandwidth based on the slice request and informs the GPU server daemon of the available NIC ports per slice, which the daemon exposes to NCCL via \texttt{NCCL\_SOCKET\_IFNAME}.

\begin{figure}[h!]
  \includegraphics[width=0.45\textwidth]{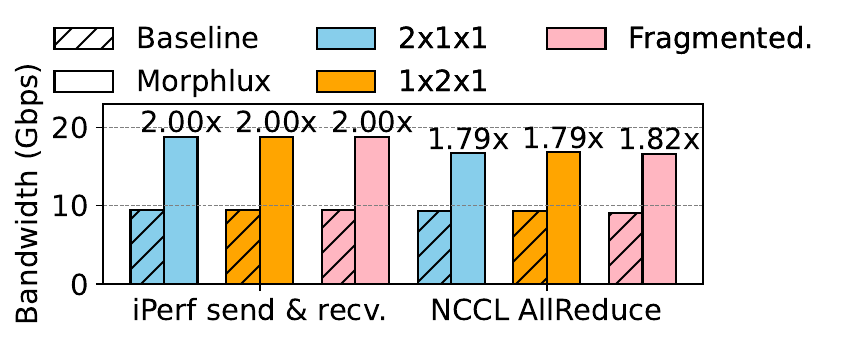}
  \caption{Bandwith improvement.}
  \label{fig:bw_improve}
  \vspace{-2mm}
\end{figure}

\begin{figure*}[h]
  \centering
  \begin{subfigure}[b]{0.33\textwidth}
    \includegraphics[width=\textwidth]{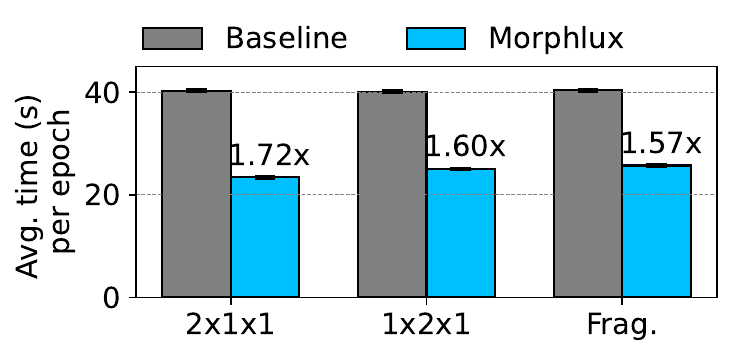}
    \caption{Workload speedup.}
    \label{fig:time_improve} 
  \end{subfigure}  
  \begin{subfigure}[b]{0.3\textwidth}
    \includegraphics[width=\textwidth]{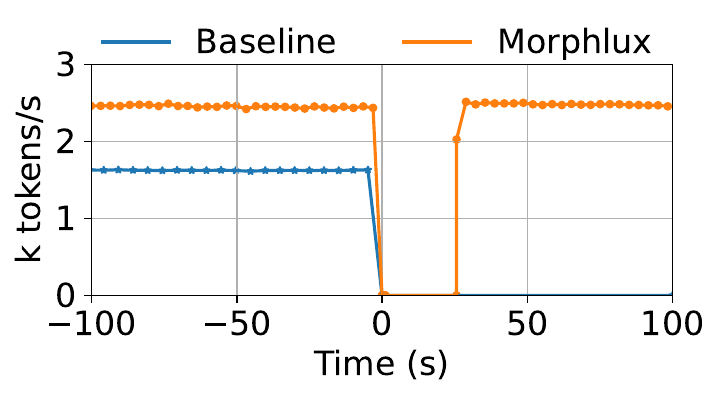}
    \caption{Fault tolerance experiment.}
    \label{fig:fault_tolerance}
  \end{subfigure}
  \begin{subfigure}[b]{0.34\textwidth}
    \includegraphics[width=\textwidth]{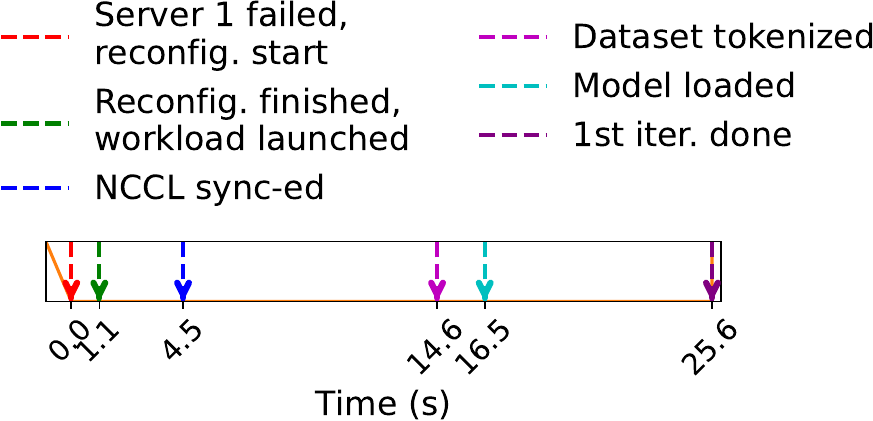}
    \caption{Zoomed-in view of Figure \ref{fig:fault_tolerance}.}
    \label{fig:fault_tolerance_zoomed_in}
  \end{subfigure}
  \vspace{4mm}
  \caption{\ref{fig:time_improve} shows \sysname accelerates the workload. \ref{fig:fault_tolerance} and \ref{fig:fault_tolerance_zoomed_in} (zoomed-in view of failure in \ref{fig:fault_tolerance}) show \sysname's fault handling and recovery . Workload: Llama-3.2-1B full parameter fine-tuning, wikitext-103 dataset, batch/GPU 8, DDP, 2 workers.}
  
    \label{fig:hardwareresults}
\end{figure*}
\vspace{-2mm}
\subsection{Improved bandwidth and compute utilization}
We allocate slices of size 2 with different shapes, $2\times1\times1$ and $1\times2\times1$, and execute workloads. In the baseline (Figure~\ref{fig:hardware_topo} (i)), one of the two NIC ports on each GPU server remains unused due to static partitioning. \sysname leverages optical reconfiguration to redirect this bandwidth along the communication group within the slice (Figures~\ref{fig:hardware_topo} (ii), (iii)). Figure~\ref{fig:bw_improve} demonstrates that \sysname increases effective bandwidth between accelerators by utilizing spare ports, and this improvement applies to any slice on the fabric.

Figure~\ref{fig:bw_improve} shows that \sysname increases the iperf bandwidth between XPUs by $2\times$ and \allreduce bandwidth by $1.8\times$. Figure~\ref{fig:time_improve} shows that the improvement in bandwidth directly reduces the end-to-end finetuning iteration time of \texttt{Llama-3.2-1B} by $1.72\times$. This speedup is due to the lower \allreduce time during gradient synchronization, resulting from the 2$\times$ bandwidth improvement of \sysname.

\begin{figure}
    \begin{center}
        \includegraphics[width=0.36\textwidth]{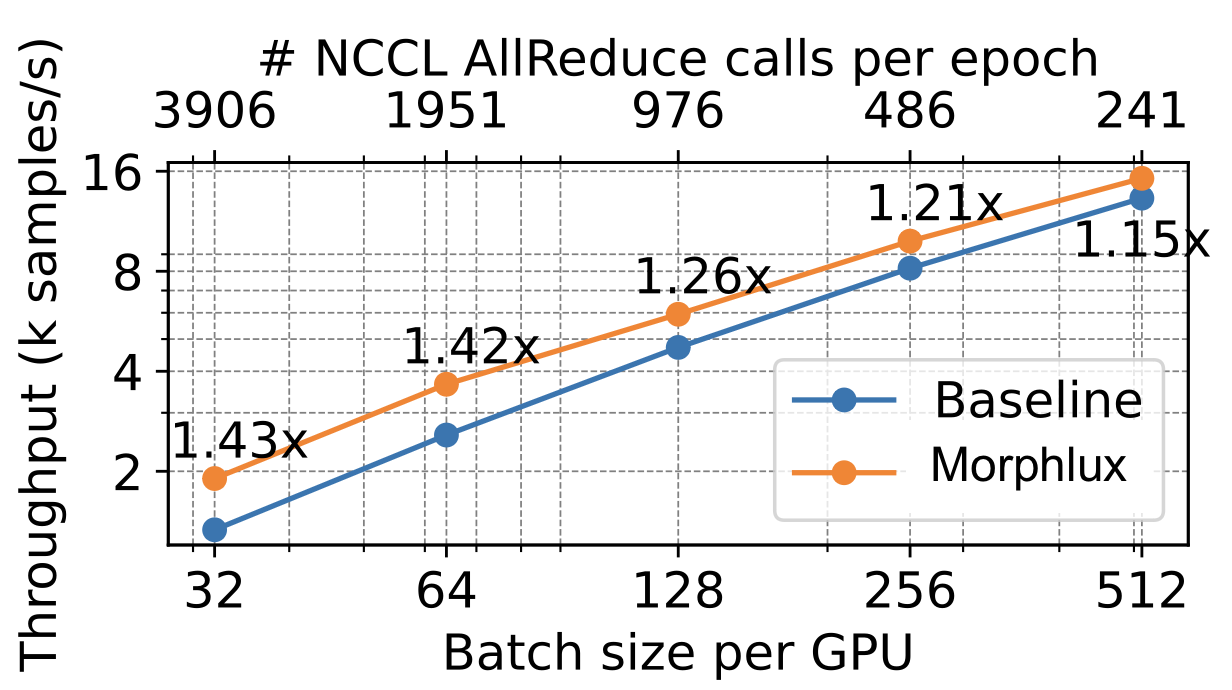}
    \end{center}
    \caption{Training throughput of Resnet50, dataset size 50k, DDP.}
    \label{fig:throughput_improve} 
    \vspace{-2mm}
\end{figure}

\myparab{Benefit of Improved Bandwidth.}
Using our hardware testbed, we show the benefit of \sysname's bandwidth redirection on training throughput (thousand samples/second) of \texttt{Resnet50} \cite{he2016deep} in Figure \ref{fig:throughput_improve}. The throughput improvement becomes more prominent when the number of \texttt{AllReduce} calls increases per epoch as batch size decreases. The speedup of LLM fine-tuning workload enabled by \sysname is also observed across different batch sizes, as shown in Table \ref{tab:llama_ft}. \sysname bandwidth redirection leads to at most 1.72$\times$ speedup for \texttt{Llama3.2-1B} fine-tuning. Larger batch size contributes to more efficient computation and thus lower latency.

\begin{table}[h!]
  \centering
  \resizebox{\columnwidth}{!}{%
  \begin{tabular}{c c c c c}
     \toprule
      Batch size & Topology & Avg. epoch time (s) & Relative speedup ($\times$)\\ \midrule
      \multirow{2}{*}{2 per GPU} & Baeline & 144.184 $\pm$ 0.175 & 1 \\
                                  & Morphlux   & 89.739 $\pm$ 0.789 & 1.61   \\ \hline
      \multirow{2}{*}{4 per GPU} & Baseline   & 75.222 $\pm$ 0.261 & 1  \\
                                  & Morphlux   & 46.361 $\pm$ 0.086 & 1.62  \\ \hline
      \multirow{2}{*}{8 per GPU} & Baseline  & 40.249 $\pm$ 0.257 & 1  \\
                                  & Morphlux & 23.369 $\pm$ 0.236 & 1.72  \\
      \bottomrule
  \end{tabular}}
  \vspace{5mm}
  \caption{Llama-3.2-1B full parameter fine-tuning, wikitext-103 dataset (128 samples selected), batch/GPU 8, DDP, 2 workers.}
  \label{tab:llama_ft}
\end{table}



\myparab{Fragmented allocations.} With \sysname's fragmented slice allocation, GPUs 0 and 1, which would remain idle under regular shape allocation, can be utilized by slice 1 (Figure~\ref{fig:hardware_topo}(iv)). Figures~\ref{fig:bw_improve} and \ref{fig:time_improve} show that \sysname improves bandwidth by $2\times$ and finetuning iteration time by $1.6\times$ with fragmented slices because \sysname not only connects arbitrary servers with optical circuits but also redirects bandwidth along these circuits. The performance gains are identical to regular slices due to the homogeneous networking properties of the photonic fabric, which ensures similar latency and bandwidth across different circuits.




\subsection{Fault tolerance}

To evaluate the fault tolerance of \sysname, we simulate a hardware failure during workload execution. In this experiment, we reserve server 3 to tolerate faults. We run the \texttt{Llama-3.2-1B} fine-tuning job on two servers (server 0 and server 1) and terminate the process on server 1. The \sysname\ fault manager detects the failure and reconfigures the fabric to establish circuits between server 0 and server 3, enabling the workload to resume execution (Figure~\ref{fig:hardware_topo}(v)).  

Figure~\ref{fig:fault_tolerance} and \ref{fig:fault_tolerance_zoomed_in} present the fault-handling timeline. The baseline suffers from 33.53\% lower throughput on average due to bandwidth underutilization and ultimately fails to complete, as no healthy accelerators remain to form a regular slice for migration. In contrast, \sysname\ enables continued execution by reallocating the workload to a fragmented slice. The fabric reconfiguration overhead is minimal (4.58\%) compared to the cost of restarting the workload. Note that \sysname took roughly 1 second to reprogram the photonic mesh and create circuits to \emph{in-place patch in} the healthy GPU. Bulk of the reaction time in Figure~\ref{fig:fault_tolerance_zoomed_in} is due to software delays like moving data, model and synchronizing NCCL.

\section{Large-scale evaluation of \sysname}
\label{sec:sim}

We evaluate the benefits of \sysname at scale using our novel TPU cluster simulator. We sample slice sizes and shapes from the TPU cluster slice distribution~\cite{googletpuv4}. We target slice sizes smaller than a rack (4, 8, 16 and 32) since bandwidth is underutilized and resources are fragmented by slices when they are smaller than a rack~(\S\ref{subsec:bw_underutil}).

\myparab{End-to-end simulation.}  Tenants fine-tune pre-trained models~\cite{roberta,scaling_finetuning} using datasets of only a few thousand samples. Empirical studies show that fine-tuning transformers on small datasets achieves optimal accuracy with small batch sizes of 16-64~\cite{roberta,llama2,scaling_finetuning}. We use the transformer model in state-of-the-art training simulator FlexNet from prior work~\cite{topoopt} for simulations. We set the hidden dimension to match Meta Llama's 4096 and evaluate with batch sizes of 8, 16, 32, 64 on slices with 4, 8, 16, 32 TPUs respectively. We describe the communication groups and collective algorithms we use during the simulations in Appendix~\ref{appendix:flexnet}.


\begin{figure*}[h!]
    \begin{subfigure}{0.27\textwidth}
        \centering
        \includegraphics[width=\textwidth]{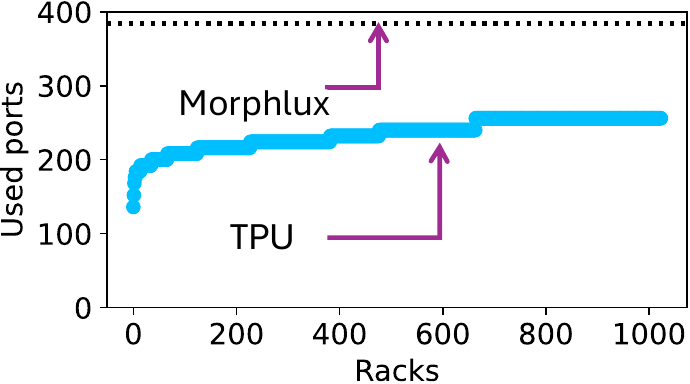}
        \caption{Ports usage}
        \label{fig:port_usage}
    \end{subfigure}
    \begin{subfigure}{0.3\textwidth}
        \centering
        \includegraphics[width=\textwidth]{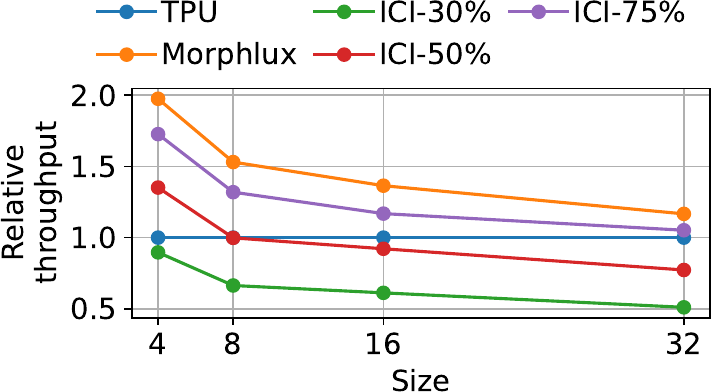}
        \caption{Bert throughput}
        \label{fig:bert_improvement}
    \end{subfigure}
    \begin{subfigure}{0.3\textwidth}
        \centering
        \includegraphics[width=\textwidth]{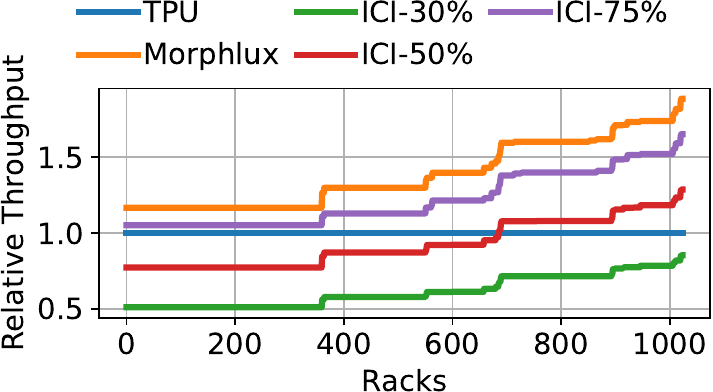}
        \caption{Rack throughput}
        \label{fig:rack_throughput}
    \end{subfigure}
    \vspace{3mm}
    \caption{~\ref{fig:port_usage} shows how many of the available ports are allocated to slices. ~\ref{fig:bert_improvement} shows the improvement in transformer training throughput. ~\ref{fig:rack_throughput} shows \sysname improves the overall throughput of a rack by up-to 60\% for finetuning.}
    \label{fig:bw_underutil_fix}
\end{figure*}

\subsection{Bandwidth utilization}
\label{subsec:bw_util_eval}
We measure bandwidth utilization in terms of the number of SerDes ports in a rack that are usable by the slices without causing contention. We sample slices from the production distribution and allocate it until the cluster is fully occupied. Each cluster has 64 racks. To fully understand \sysname's performance in large-scale, we repeat this experiment 16 times with a new cluster to have ($16\times64$) 1024 rack states. We compare the bandwidth utilization of an optically interconnected that uses \sysname with the following baselines.

We built a best effort allocator to assign TPU cluster resources to slice requests. Our slice allocator finds a contiguous set of TPUs in the desired dimensions by searching the racks sequentially. This baseline models the default behavior of the TPU cluster where the bandwidth of some ports on a TPU are unused if they are connected to links shared by multiple slices (\S\ref{subsec:bw_underutil}). TPUs are equipped with ICI switches that can route packets. Using ICI switches to redirect bandwidth from the unused ports causes contention on links shared by multiple slices and reduces the observed bandwidth. We model this by using all ports on a TPU for communication within a slice but reduce the observed bandwidth of each port by different amounts (70\%, 50\%, 25\%) to in case of contention.

\myparab{Underutilization of ports.} In Figure~\ref{fig:port_usage}, the blue line shows that up to 50\% of ports in electrical racks remain unused by slices, with 1024 points representing the simulated racks. In contrast, \sysname reallocates bandwidth from unused ports to active dimensions, achieving 100\% utilization. Bandwidth utilization varies across racks based on slice distribution within each rack.

\myparab{Impact on finetuning.} Figure~\ref{fig:bert_improvement} shows that \sysname improves the finetuning throughput by up-to $2\times$. The improvement is higher for smaller sizes because they incur more bandwidth underutilization in TPU slices. ICI-X\% baselines utilize X\% of the total bandwidth on a TPU with ICI routing and are consistently worse than \sysname which utilizes 100\% of the bandwidth. At larger slices, the ICI-50\% and 30\% slices perform worse than regular TPU slices because the impact of contention on collective communication is higher than the impact of bandwidth underutilization in regular TPU slices. In Figure~\ref{fig:rack_throughput} we report the sum of throughput of various slices allocated in the rack. \sysname improves the overall throughput of a rack significantly, 50\% of the racks have more than 30\% improvement in their throughput.

\subsection{Allocating fragmented racks}

\sysname allocates larger slices even when contiguous servers are unavailable by connecting discontiguous free servers via optical circuits. We begin by evenly distributing all 4096 TPUs in the cluster across slices of size 4, 8, 16, and 32, allocating 1024 TPUs to each. Next, we randomly deallocate slices until 30\% of the TPUs are free. From this state, we evaluate using two distributions: (1) allocating only 32-TPU slices until available TPUs are exhausted and (2) allocating 16 and 32 TPU slices which we denote by (16, 32). Note that fragmentation prevents allocation only when the size of slice allocation requests is larger than the contiguous free slices within a rack. Thus, we deallocate slices of all sizes but subsequently allocate only larger slices (16 and 32).

\begin{figure}[h!]
    \begin{subfigure}{0.25\textwidth}
        \centering
        \includegraphics[width=\textwidth]{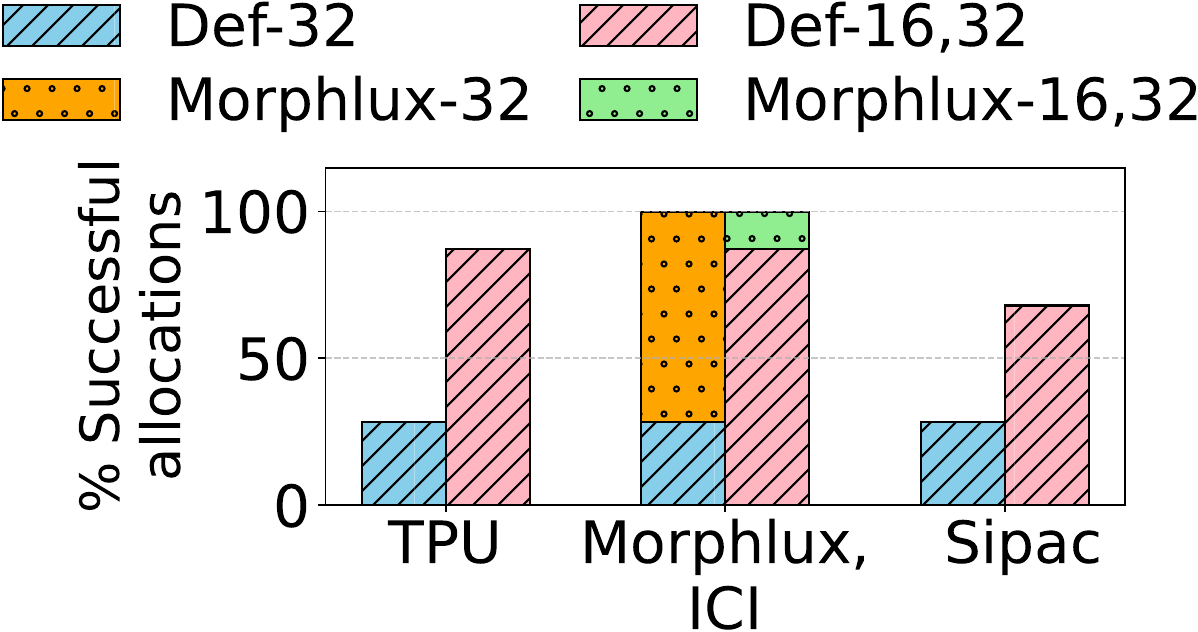}
        \caption{Fragmented allocations}
        \label{fig:fragmentation_success}
    \end{subfigure}
    \begin{subfigure}{0.22\textwidth}
        \centering
        \includegraphics[width=\textwidth]{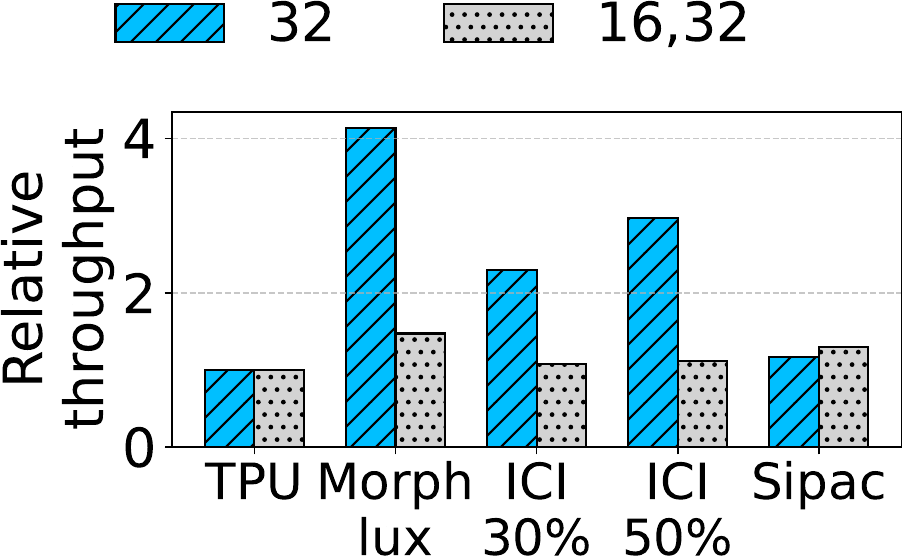}
        \caption{Transformer throughput}
        \label{fig:impact_on_throughput}
    \end{subfigure}
    \vspace{1mm}
    \caption{Figure~\ref{fig:fragmentation_success} shows the fraction of new allocations that require fragmented allocator. Figure~\ref{fig:impact_on_throughput} shows the relative throughput of larger slices allocated using \sysname's fragmented allocator with respect to default allocator.}
    \label{fig:fragmentation_fix}
    \vspace{-2mm}
\end{figure}

\myparab{Baselines.} We again use the TPU and ICI baselines. Here, we use ICI switching to not only utilize additional ports but also to route data between fragmented servers. We also compare \sysname with another photonics interconnect, SiPAC~\cite{karen2,karen_journal} which uses the Bcube topology. We implement their allocation policy~\cite{karen_journal} that maps slices sequentially to accelerators in the topology and can satisfy a slice request only if there are free contiguous accelerators.

\myparab{Fragmented allocations.} Figure~\ref{fig:fragmentation_success} shows that default allocators in TPU and SiPAC fail to find contiguous resources for approximately 75\% of slice requests in the 32-TPU distribution. \sysname overcomes this limitation by leveraging optical interconnects to establish logical topologies from physically discontiguous TPUs, successfully allocating all 75\% of previously unallocated requests. In the (16,32) TPU distribution, the default allocators perform better due to (1) a reduced number of 32-TPU slice requests and (2) a higher probability of finding 16 contiguous TPUs compared to 32. \sysname achieves allocation performance equivalent to an ideal switch with all-to-all TPU connectivity, ensuring allocation as long as sufficient TPUs are available in the rack.

\myparab{Impact on finetuning.} Figure~\ref{fig:impact_on_throughput} shows that \sysname improves the overall throughput across all the new slice requests by $4\times$ and $1.3\times$ in 32 TPUs and (16,32) TPUs distributions because \sysname allocates 3 times more 32 TPUs slices and 1.2 times more (16,32) TPUs slices and slices in \sysname also have higher bandwidth enabled by bandwidth redirection. Although SiPAC supports bandwidth redirection, which we model in our baseline, its performance is marginally better than the TPU baseline because SiPAC's allocator satisfies similar number of new slice requests as TPU in both the distributions and the benefit of bandwidth redirection is lower at larger slices like 32 and 16 TPUs. ICI switching baseline's performance is at least 40\% lower than \sysname while using 50\% of the total TPU bandwidth. The ILP associated with fragmented allocator converges in less than 600 milliseconds in all of our experiments and finds optical circuit routes that do not require additional fibers.


\begin{figure}[h!]
    \centering
    \includegraphics[width=0.25\textwidth]{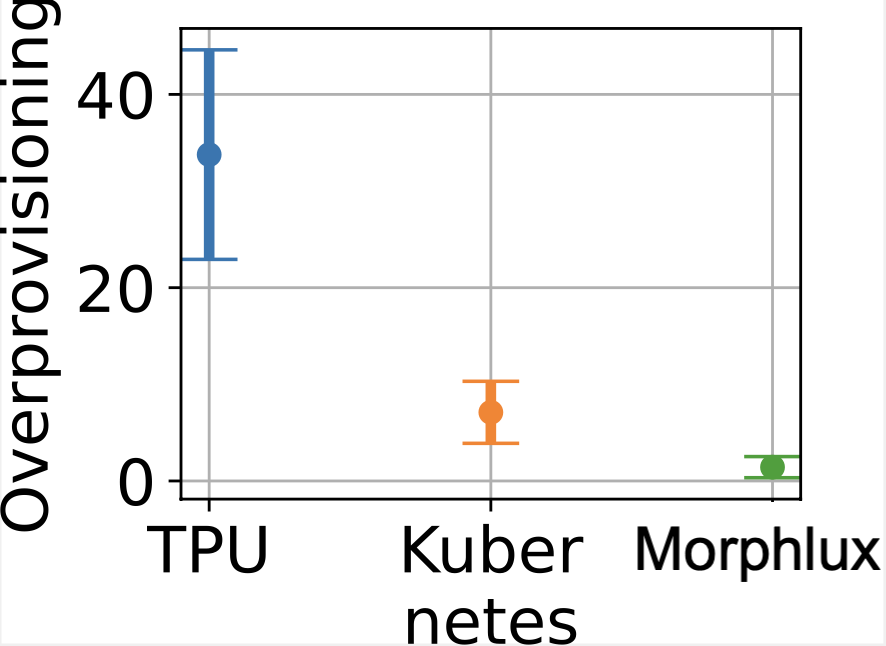}
    \caption{shows the average overprovisoning over 1024 racks with standard deviation as error bars.}
    \label{fig:fault_overprovison}
    \vspace{-4mm}
\end{figure}

\subsection{Fault tolerance at chip granularity}
\label{subsec:fault_sim_eval}
In this experiment, we fully allocate the cluster in our simulator and then randomly fail 1 to 4 chips in each rack. In \sysname, we reserve 4 TPUs, \ie one full server at (0, 0, 0) in each rack to handle faults within the rack. We compare \sysname with the job migration policy of TPU clusters~\cite{tpu-resilience}, which migrates the entire job, even if a single TPU fails. Additionally, we evaluate Kubernetes, a cluster management framework, using a recent policy~\cite{megascale} that evicts faulty-chip servers during training and replaces them with free servers in a CLOS-like topology. 

\myparab{Hardware underutilization.} In Figure~\ref{fig:fault_overprovison}, we simulate TPU failures across 1024 racks in 16 clusters and measure the TPUs required to handle faults using \sysname{} and baseline approaches. We define overprovisioning as the excess TPUs needed beyond the number of failed TPUs. For example, if 2 TPUs fail in a 32-TPU slice, the TPU baseline requires 32 replacements, resulting in an overprovisioning of $32 - 2 = 30$. Figure~\ref{fig:fault_overprovison} reports the average overprovisioning with standard deviation as error bars. \sysname reduces TPU requirements by an order of magnitude compared to the TPU baseline, by $3\times$ compared to Kubernetes, and matches the ideal switch by establishing optical circuits between neighbors of failed TPUs and reserved TPUs.

\section{Related work}
\label{sec:discussion}

Recent work uses optical components for failure resilience~\cite{arrow}, capacity augmentation~\cite{radwan,shoofly} and bulk data transfers~\cite{owan}. Recently, there has been growing commercial interest in making optics an active part of routing in datacenter interconnects. For instance, Google has recently replaced the spine layer packet switches in their datacenter interconnects with optical circuit switches~\cite{googletpuv4,jupiter-rising,jupiter-evolving}. Researchers have used optics for reconfiguring the datacenter interconnects~\cite{topoopt,rotornet,firefly,karen4}. Researchers have used silicon photonic datacenter fabrics for improving ML training~\cite{karen1,karen2,karen3, sipml, karen-ofc23}. While optical connectivity has permeated datacenter racks, interconnects between chips on compute servers remain electrical. \sysname addresses this gap.





\label{endOfBody}

\bibliographystyle{ACM-Reference-Format}
\bibliography{reference, reference-old, more-refs}
\appendix

\section{Bandwidth Underutilization examples}
\label{appendix:examples}

\begin{figure}[h]
    \centering 
    \begin{subfigure}[b]{0.45\textwidth}
        \includegraphics[width=\textwidth]{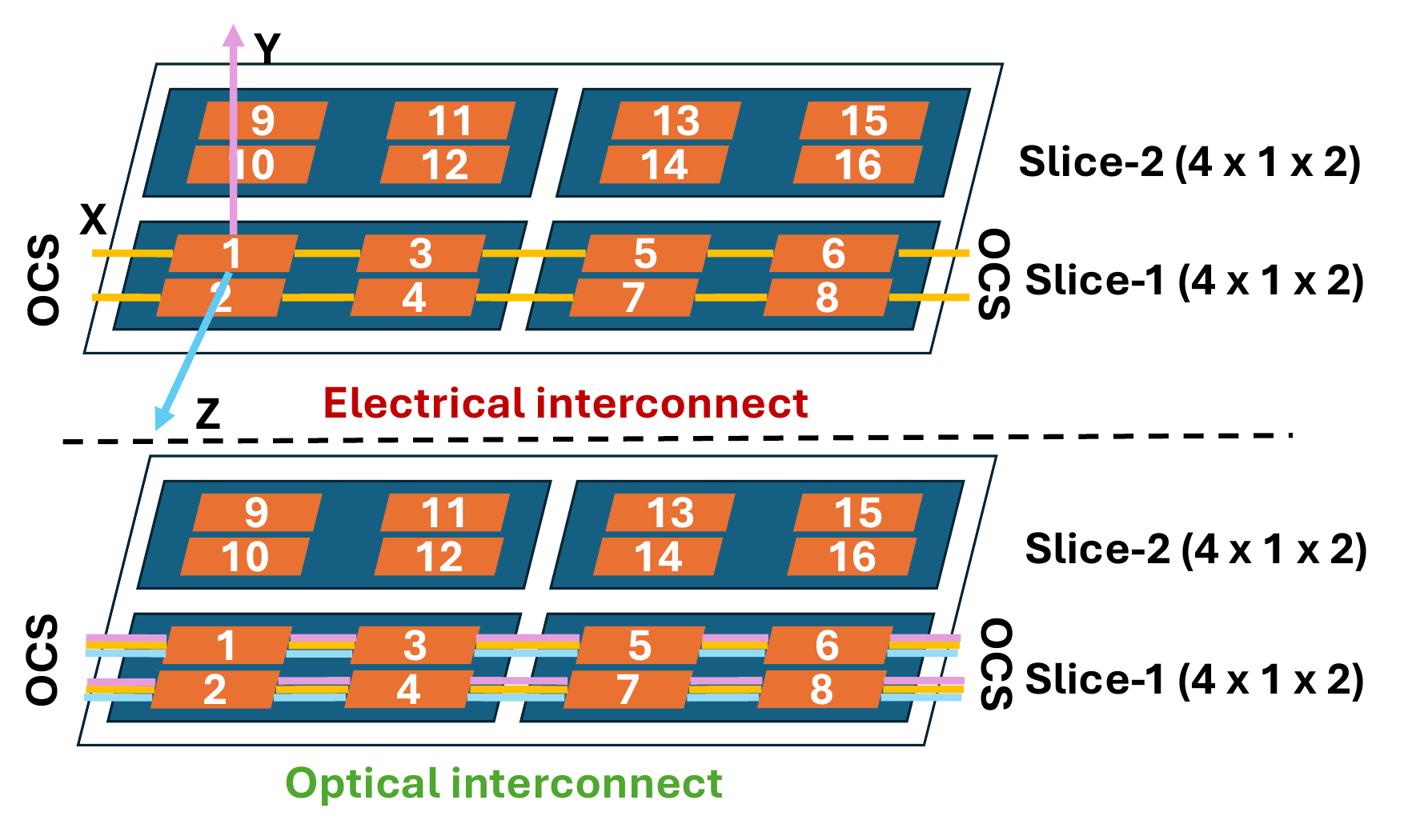}
    \end{subfigure}
    \vspace{0.4cm}
    \caption{\small{
    shows a rack with multiple small slices.}}
    \label{fig:tpus}
\end{figure}

A slice consists of a subset of TPU chips allocated to a 
single cloud tenant. Typically, slices can only be allocated 
in regular shapes, forming tori of specific dimensions~\cite{tpu-allocate}.
Tenants deploy their training and inference jobs on the allocated 
TPU slice, during which collective communication primitives 
are executed over the slice torus using the multi-dimensional bucket algorithm~\cite{bucket-ring}.
We note that TPU slices allocated to customers do not always span
multiple racks. Most inference workloads need smaller slices as 
model sizes rarely exceed the memory of all TPUs in a rack.

The multi-dimensional bucket algorithm sequentially executes data 
transfers in a ring across all the dimensions of the torus. So, in a 3D 
torus, 3 rings are executed in the order of dimensions $XYZ$. As a 
result, connectivity in two of the three dimensions is always 
underutilized since only one ring is active at a given time. 
Intermediate computation prevents overlapping the three rings 
to prevent the interconnect under-utilization. To address this
under-utilization, researchers have proposed algorithms that subdivide the
data buffer to execute several multidimensional bucket algorithms 
in different sequences of dimensions, \eg $YZX$ and $ZXY$,
simultaneously such that all the dimensions are utilized throughout 
the collective~\cite{swing}. However, this does not offer 
better performance, as we will discuss later.

We define congestion in a direct-connect topology as 
the scenario where multiple transfers occur simultaneously 
on the same link, similar to the definition in CS theory~\cite{makespan1,makespan2}.
A slice optimally utilizes all 3 dimensions only when there is no 
congestion in any dimension. Note that due to the design of a torus, 
this can only happen when a slice spans multiple racks. For instance, consider 
Figure~\ref{fig:tpus} with multiple slices. All the slices share 
rings along the Y dimension. If we avoid the Y dimensional ring, then
the bandwidth is underutilized by 33\% because the slices have access
to only 2 of the 3 dimensions. Figure \ref{fig:tpus} demonstrates 
how even smaller slices can suffer up-to 66\% lower bandwidth. 
Slice 1 and 2 share both the Y and Z dimensions with other slices
and can only execute the X dimensional ring without causing
congestion, which leads to suboptimal performance.

\begin{table}[h]
    \centering
    \resizebox{\columnwidth}{!}{%
    \begin{tabular}{lcccc}
        \toprule
        \textbf{Elec. $\alpha$ cost} & \textbf{Optics $\alpha$ cost} & \textbf{Elec. $\beta$ cost} & \textbf{Optics $\beta$ cost} \\
        \midrule
        $7 \times \alpha$ & $7 \times \alpha + r$ & $\frac{N}{1} \cdot \left(\frac{8 - 1}{8}\right) \cdot (3 \times \beta)$ & $\frac{N}{1} \cdot \left(\frac{8 - 1}{8}\right) \cdot (1 \times \beta)$ \\
        \bottomrule
    \end{tabular}
    }
    \vspace{0.3cm}
    \caption{\reducescatter costs of Slice-1.}
    \vspace{-1em}
    \label{tab:slice1_costs}
\end{table}

\myparab{Impact of the underutilized bandwidth on $\alpha-\beta$ costs.}
We use the $\alpha-\beta$ cost model~\cite{taccl} to reason about
the cost of collective communication.
$\alpha$ cost is the software overhead of sending data buffers.
$\beta$ cost is the transmission delay, which is inversely 
proportional to link bandwidth. The \allreduce bucket ring algorithm on a 
$D$ dimensional torus has $D$ \reducescatter operations followed by 
$D$ \allgather operations and can reach optimal $\beta$ cost with simultaneous 
rings in all $D$ dimensions. However, Table~\ref{tab:slice1_costs}
shows that the cost of one \reducescatter on Slice-1 is 
proportional to $3$ times the optimal ($\frac{1}{bandwidth}$), because 
the slice is utilizing the bandwidth of only one dimension of the torus.

\begin{algorithm}[t]
    \textbf{Inputs:}\\
    \begin{tabular}{rl}
        $G': \langle L, T \rangle$: & Slice request graph with $L$ slots \\
        & and $T$ edges. \\
        $G: \langle S, I \rangle $: & Physical rack's graph with $S$ servers \\
        & and $I$ edges. \\
        ${P}(u, v)$: & Set of paths between servers $u$ and $v$, \\
                    & $\forall u, v \in {S}$, where each $Q \in {P}(u, v)$ \\
                    & is a list of edges $i \in {I}$ \\[1em]
    \end{tabular}
    
    \textbf{Outputs:}\\
    \begin{tabular}{rl}
        $x_{a,b}$: & $\in \{0, 1\}$, binary variable indicating if server \\
                  & $a \in {S}$ is mapped to slot $b \in {L}$ \\
        $r_{i,u,v}$: & $\in \{0, 1\}$, binary variable indicating if $i^{th}$ path \\
                    & $\in {P}(u, v)$ is selected for communication \\
                    & between servers $u$ and $v$ \\
        $z$: & $\in \mathbb{Z}$, maximum edge overlap \\[1em]
    \end{tabular}
    
    \textbf{Minimize:} $z$ \\
    for all edges in the rack, $\forall{e \in I}$ \\
    for all routes $Q$ passing through this edge $Q \subset P$ \\
    $z \geq \sum_{i \in Q}$ $r_{i,u,v} \cdot 4 + b(e)$ \\
    where $b(e)$ is the number of existing circuits on edge $e$ \\[1mm]

    \textbf{subject to} \\[1mm]
    $\sum_{a \in {L}} x_{a,b} \leq 1, \text{ } \forall b \in {F}$ \\
    $\sum_{b \in {F}} x_{a,b} = 1, \text{ } \forall a \in {L}$ \\

    $\forall m, n \in S, \text{ } \forall a, b \in L, \text{ } x_{a,m} \cdot x_{b,n} = 1, \text{ } e_{a,b} \in T \implies $
    \\ 
    $\sum_{i \in P(m, n)} r_{i,m,n} = 1$

    
    \caption{Fragmented slice allocator.}
    \label{alg:fragmented_allocator}
  \end{algorithm}

\section{Implementation}
\label{appendix:implementation}
We implement \orchestrator with a TPU cluster simulator and a hardware testbed which we evaluate in \S~\ref{sec:hardwareeval}.

\subsection{TPU cluster simulator implementation}
We have implemented a simulator to model the TPU cluster described in~\cite{googletpuv4,tpu-resilience}. We implement the simulator in Python and is available in the source code repository that we have released. We use Python classes to model a TPU, OCS, Rack (block) and Slices. The simulator implements \orchestrator by maintaining cluster state and offers methods to allocate and deallocate slices. During the default allocation phase, the simulator searches through the blocks sequentially for contiguously available TPUs to satisfy the allocation request. Upon failure and if fragmented allocation is enabled, the simulator leverages fragmented allocator described in Algorithm~\ref{alg:fragmented_allocator}. We have implemented Algorithm~\ref{alg:fragmented_allocator} in Gurobi and integrated it with the TPU cluster simulator. The simulator also offers a method to reserve servers to manage failures and never allocates this to any slice request so that the fault tolerance policies can leverage the TPUs on the server when needed.

\subsection{\orchestrator for hardware testbed}
We implement \orchestrator in Python. In \orchestrator's allocator, we construct virtual classes to represent compute and network resources (accelerators, silicon photonic fabrics, optical fibers, \etc). Upon slice request, the allocator executes regular slice allocation logic by default. If regular slice allocation is not possible, to prevent resource fragmentation and maximize resource utilization, the allocator runs the fragmented allocation algorithm (Algorithm \ref{alg:fragmented_allocator}) and uses Gurobi \cite{gurobi} to solve the ILP problem. The fault manager uses accelerator monitoring services like NVIDIA's \texttt{nvidia-smi} and Linux network utility tools like \texttt{Ping} for resource health monitoring, and keeps track of resource status. If the fault manager detects resource failure in an active slice, it identifies healthy, available resources and generates a new allocation plan. The hardware control plane is responsible for translating logical slice configurations into physical configurations. This process involves mapping logical network topologies onto physical interconnects, which is hardware dependent. For our hardware testbed, we use the algorithm in \cite{ding2025pipswitchcircuitswitchusing} to implement the control plane.

\subsection{Hardware Control Plane Bandwidth Translation}
\label{appendix:bw_translation}
The \allocator and \faultmanager output a set of servers and paths between them that form the desired slice topology to the \controlplane. The \controlplane then starts implementing the logical topologies by assigning SerDes ports to communication groups within a slice. The \controlplane executes the following steps sequentially:
\begin{enumerate}
  \item Distribute the available ports evenly across communication groups on every fabric based on how many groups occupy that fabric.
  \item Determine the exact number of ports for each group by finding the minimum number of assigned ports to the group across all fabrics from step (1) such that the number is consistent across all fabrics occupied by the communication group.
  \item Select the rank of ports for each communication group on every fabric such that the path finding process could succeed (being able to a feasible solution).
\end{enumerate}

\section{FlexNet setup details}
\label{appendix:flexnet}
FlexNet generates an optimal parallelization strategy and its dependency graph with collective communication vertices. We simulate training iteration times using this graph under varying communication costs. FlexNet identifies collective communication groups and buffer sizes using which we partition bandwidth in \sysname's \controlplane similar to prior work~\cite{topoopt}. Within each group, we apply the Bucket algorithm~\cite{bucket-ring}, optimal for torus topologies in electrical interconnects. In \sysname, we use the Ring algorithm~\cite{topoopt,taccl}, as each node communicates with same two neighbors every step, requiring topology reconfiguration only at job initiation. Collective communication time is estimated using the $\alpha$-$\beta$ cost model~\cite{taccl}, with $\alpha$ from prior work~\cite{taccl} and $\beta$ set to TPU's ICI bandwidth of 50 GB/s~\cite{googletpuv4}. Both of these algorithms achieve identical bandwidth cost.
\end{document}